\documentclass{article} 
\usepackage{arxiv}

\usepackage[english]{babel} 
\usepackage{amssymb}
\usepackage{amsmath}
\usepackage{txfonts}
\usepackage{mathdots}
\usepackage[classicReIm]{kpfonts}
\usepackage[dvips]{graphicx} 

\usepackage{subfigure}

\usepackage[utf8]{inputenc}
\usepackage{algorithm} 
\usepackage{algpseudocode}

\usepackage{tikz}
\usetikzlibrary{backgrounds,fit,decorations.pathreplacing, shapes}  
\usepackage{tkz-graph}

\usepackage[super]{nth}
\usepackage{adjustbox}
\usepackage{nameref}

\tikzstyle{vertex}=[circle,draw=black, fill=white,sloped,minimum size=17pt,inner sep=5pt]

\usepackage{tikz}
\usetikzlibrary{decorations.pathreplacing,calc}

\usepackage{enumerate}
\usepackage{amsmath}
\usepackage{graphicx}
\usepackage{wrapfig}
\usepackage{lscape}
\usepackage{rotating}
\usepackage{epstopdf}

\usepackage{caption}
\usepackage{url}
\usepackage{hyperref}
\usepackage{float}
\usepackage[super]{nth}
\usepackage[labelfont=bf]{caption}
\usepackage[labelsep=period]{caption}

\usepackage{pgfplots} 

\pgfplotsset{compat = newest}

\newcommand{\figref}[1]%
{Figure \ref{#1}%
}

\newcommand{\tableref}[1]%
{Table \ref{#1}%
}

\newcommand{\algorithmref}[1]%
{Algorithm \ref{#1}%
}

\newcommand{\sectionref}[1]%
{Section \ref{#1}%
}

\newcommand{\lineref}[1]%
{Line \ref{#1}%
}

\algnewcommand{\LineComment}[1]{\State \(\triangleright\) #1}

\usepackage{newunicodechar}
\newunicodechar{−}{-}

\title{An Overview of Statistical Data Analysis}

\author{Rui Portocarrero Sarmento \\
LIAAD-INESC TEC \\
PRODEI - FEUP, University of Porto \\
mail@ruisarmento.com \And
Vera Costa \\
FEUP, University of Porto \\
veracosta@fe.up.pt}

\begin{document}


\maketitle

\begin{abstract}

The use of statistical software in academia and enterprises has been evolving over the last years. More often than not, students, professors, workers and users in general have all had, at some point, exposure to statistical software. Sometimes, difficulties are felt when dealing with such type of software. Very few persons have theoretical knowledge to clearly understand software configurations or settings, and sometimes even the presented results. Very often, the users are required by academies or enterprises to present reports, without the time to explore or understand the results or tasks required to do a optimal preparation of data or software settings. In this work, we present a statistical overview of some theoretical concepts, to provide a fast access to some concepts.

\end{abstract}


\section{Introduction}

Statistics is a set of methods used to analyze data. The statistic is present in all areas of science involving the collection, handling and sorting of data, given the insight of a particular phenomenon and the possibility that, from that knowledge, inferring possible new results. One of the goals with statistics is to extract information from data to get a better understanding of the situations they represent. Thus, the statistics can be thought of as the science of learning from data.

Currently, the high competitiveness in search technologies and markets has caused a constant race for the information. This is a growing and irreversible trend. Learning from data is one of the most critical challenges of the information age in which we live. In general, we can say that statistic based on the theory of probability, provides techniques and methods for data analysis, which help the decision-making process in various problems where there is uncertainty.
 
This paper presents all theory concepts that support the developed programming code, either with R or Python language, presented in \citep{sarmento2017comparative}\footnote{\url{https://bit.ly/2TRREw9}}.

\paragraph{Variables, Population, AND Samples}

 In statistical analysis, ``variable'' is the common characteristic of all elements of the sample or population to which is possible to attribute a number or category. The values of the variables vary from element to element.

\subsection{Types of variables}

 Statistical variables can be classified as categorical variables or numerical variables.

\textbf{\textit{Categorical variables}} have values that describe a ``quality'' or ``characteristic'' of a data unit, like ``which type'' or ``which category''. Categorical variables fall into mutually exclusive (in one category or another) and exhaustive (include several possible options) categories. Therefore, categorical variables are qualitative variables and tend to be represented by a non-numeric value. Categorical variables may be further described as (Mar\^{o}co, 2011):

\begin{enumerate}
\item  Nominal: the data consist of categories only.  The variables are measured in discrete classes, and it is not possible to establish any qualification or ordering. Standard mathematical operations (addition, subtraction, multiplication, and division) are not defined when applied to this type of variable. Gender (male or female) and colors (blue, red or green) are two examples of nominal variables.

\item  Ordinal: the data consist of categories that can be arranged in some exact order according to their relative size or quality, but cannot be quantified. Standard mathematical operations (addition, subtraction, multiplication, and division) are not defined when applied to this type of variable. For example, social class (upper, middle and lower) and education (elementary, medium and high) are two examples of ordinal variables. Likert scales (1-``Strongly Disagree'', 2-``Disagree'', 3-``Undecided'', 4-``Agree'', 5-``Strongly Agree'') are ordinal scales commonly used in social sciences.
\end{enumerate}

 \textbf{\textit{Numerical variables}} have values that describe a measurable quantity as a number, like ``how many'' or ``how much''. Therefore, numeric variables are quantitative variables. Numeric variables may be further described as:

\begin{enumerate}
\item  Discrete: the data is numerical. Observations can take a value based on a count of a set of distinct integer values. A discrete variable cannot take the value of a fraction of one value and the next closest value. The number of registered cars, the number of business locations, and the number of children in a family, all of which measured as whole units (i.e. 1, 2, or 3 cars) are some examples of discrete variables.

\item  Continuous: the data is numerical. Observations can take any value between a particular set of real numbers. The value given to one observation for a continuous variable can include values as precise as possible with the instrument of measurement. Height and time are two examples of continuous variables.
\end{enumerate}

\subsection{Population and Samples}

 The\textbf{\textit{ population}} is the total of all the individuals who have certain characteristics and are of interest to a researcher. Community college students, race-car drivers, teachers, and college-level athletes can all be considered populations. 

 It is not always convenient or possible to examine every member of an entire population. For example, it is not practical to ask all students which color they like. However, it is possible, to ask the students of three schools the preferred color. This subset of the population is called a sample.

 A \textbf{\textit{sample}} is a subset of the population. The reason for the sample's importance is because in many models of scientific research, it is impossible (from both a strategic and a resource perspective) the study of all members of a population for a research project. It just costs too much and takes too much time. Instead, a selected few participants (who make up the sample) are chosen to ensure the sample is representative of the population. And, if this happens, the results from the sample could be inferred to the population, which is precisely the purpose of inferential statistics - using information on a smaller group of participants makes possible to understand to all population.

 There are many types of samples, including a random sample, a stratified sample, and a convenience sample, but they all have the goal to accurately obtain a smaller subset of the larger set of total participants, such that the smaller subset is representative of the larger set.

 \textbf{}

\subsection{Independent and Paired Sample}

 The relationship or absence of the relationship between the elements of one or more samples defines another factor of classification of the sample, particularly important in statistical inference. If there is no type of relationship between the elements of the samples, it is called \textbf{\textit{independent samples}}. Thus, the theoretical probability of a given subject belonging to more than one sample is null. On the opposite, if the same subject composes the samples based on some unifying criteria (for example, samples in which the same variable are measured before and after specific treatment on the same subject), it is called \textbf{\textit{paired samples}}. In such samples, the subjects who are purposely tested are related. It can even be the same subject (e.g., repeated measurements) or subject with paired characteristics (in statistical blocks studies).

\section{Descriptive Analysis}

 Descriptive statistics are used to describe the essential features of the data in a study. It provides simple summaries about the sample and the measures. Together with simple graphics analysis, it forms the basis of virtually every quantitative analysis of data. Descriptive statistics allows presenting quantitative descriptions in a convenient way. In a research study, it may have lots of measures. Or it may measure a significant number of people on any measure. Descriptive statistics helps to simplify large amounts of data in a sensible way. Each descriptive statistic reduces lots of data into a simpler summary.

\paragraph{Frequency Distributions}

 FrequencyFrequency distributions are visual displays that organize and present frequency counts (n) so that the information can be interpreted more easily. Along with the frequency counts, it may include relative frequency, cumulative, and cumulative relative frequencies.

\begin{enumerate}
\item  The\textit{ frequency} (n) is the number of times a particular variable assumes that value.

\item  The \textit{cumulative frequency} (N) is the number of times a variable takes on a value less than or equal to this value.

\item  The \textit{relative frequency} (f) is the percentage of the frequency.

\item  The \textit{cumulative relative frequency} (F) is the percentage of the cumulative frequency.
\end{enumerate}

 Depending on the variable (categorical, discrete or continuous), various frequency tables can be created.

\begin{table}
\centering
\begin{tabular}{|p{1.1in}|p{0.4in}|p{0.4in}|p{0.4in}|p{0.4in}|p{0.4in}|} \hline 
List of responses:  & Blue & Red & Blue & White & Green \\ \hline 
 & White & Blue & Red & Blue & Black \\ \hline
\end{tabular}

\end{table}

\begin{table}
\centering
\begin{tabular}{|p{1.1in}|p{0.4in}|p{0.3in}|p{0.3in}|p{0.3in}|p{0.3in}|} \hline 
FrequencyFrequency Distribution: & \textbf{Color} & \textbf{n} & \textbf{N} & \textbf{f} & \textbf{F} \\ \hline 
 & Blue & 4 & 4 & 0.4 & 0.4 \\ \hline 
 & Red & 2 & 6 & 0.2 & 0.6 \\ \hline 
 & White & 2 & 8 & 0.2 & 0.8 \\ \hline 
 & Green & 1 & 9 & 0.1 & 0.9 \\ \hline 
 & Black & 1 & 10 & 0.1 & 1.0 \\ \hline 
 & \textbf{Total} & 10 &  & 1 &  \\ \hline 
\end{tabular}
\caption{Example 1:} favorite color of 10 individuals -- categorical variable
\end{table}

\begin{table}
\centering
\begin{tabular}{|p{1.1in}|p{0.3in}|p{0.3in}|p{0.3in}|p{0.3in}|p{0.3in}|p{0.3in}|p{0.3in}|p{0.3in}|p{0.3in}|p{0.3in}|} \hline 
List of responses:  & 20 & 22 & 21 & 24 & 21 & 20 & 20 & 24 & 22 & 20 \\ \hline 
 & 22 & 24 & 21 & 25 & 20 & 23 & 22 & 23 & 21 & 20 \\ \hline 
\end{tabular}

\end{table}

\begin{table}
\centering
\begin{tabular}{|p{1.1in}|p{0.4in}|p{0.3in}|p{0.3in}|p{0.3in}|p{0.3in}|} \hline 
Frequency distribution: & \textbf{Age} & \textbf{n} & \textbf{N} & \textbf{f} & \textbf{F} \\ \hline 
 & 20 & 6 & 6 & 0.3 & 0.3 \\ \hline 
 & 21 & 4 & 10 & 0.2 & 0.5 \\ \hline 
 & 22 & 4 & 14 & 0.2 & 0.7 \\ \hline 
 & 23 & 2 & 16 & 0.1 & 0.8 \\ \hline 
 & 24 & 3 & 19 & 0.15 & 0.95 \\ \hline 
 & 25 & 1 & 20 & 0.05 & 1 \\ \hline 
 & \textbf{Total} & 20 &  & 1 &  \\ \hline 
\end{tabular}
\caption{Example 2:} age of 20 individuals -- discrete numerical variable
\end{table}

\begin{table}
\centering
\begin{tabular}{|p{1.1in}|p{0.3in}|p{0.3in}|p{0.3in}|p{0.3in}|p{0.3in}|p{0.3in}|p{0.3in}|p{0.3in}|p{0.3in}|p{0.3in}|} \hline 
List of responses:  & 1.58 & 1.56 & 1.77 & 1.59 & 1.63 & 1.58 & 1.82 & 1.69 & 1.76 & 1.60 \\ \hline 
 & 1.73 & 1.51 & 1.54 & 1.61 & 1.67 & 1.72 & 1.75 & 1.55 & 1.68 & 1.65 \\ \hline 
\end{tabular}
\end{table}

\begin{table}
\centering
\begin{tabular}{|p{1.1in}|p{0.7in}|p{0.3in}|p{0.3in}|p{0.3in}|p{0.3in}|} \hline 
Frequency distribution: & \textbf{Interval} & \textbf{n} & \textbf{N} & \textbf{f} & \textbf{F} \\ \hline 
 & ]1.50, 1.55] & 3 & 3 & 0.15 & 0.15 \\ \hline 
 & ]1.55, 1.60] & 5 & 8 & 0.25 & 0.4 \\ \hline 
 & ]1.60, 1.65] & 3 & 11 & 0.15 & 0.55 \\ \hline 
 & ]1.65, 1.70] & 3 & 14 & 0.15 & 0.7 \\ \hline 
 & ]1.70, 1.75] & 3 & 17 & 0.15 & 0.85 \\ \hline 
 & ]1.75, 1.80] & 2 & 19 & 0.1 & 0.95 \\ \hline 
 & ]1.80, 1.85] & 1 & 20 & 0.05 & 1 \\ \hline 
 & \textbf{Total} & 20 &  & 1 &  \\ \hline 
\end{tabular}
\caption{Example 3:} height of 20 individuals -- continuous numerical variable
\end{table}

\paragraph{Measures of Central Tendency and Measures of Variability}

 A measure of central tendency is a numerical value that describes a data set, by attempting to provide a ``central'' or ``typical'' value of the data \citep{mccune2009practice}.  As such, measures of central tendency are sometimes called measures of central location. They are also classed as summary statistics. 

 Measures of central tendency should have the same units as those of the data values from which they are determined. If no units are specified for the data values, no units are specified for the measures of central tendency. 

 The mean (often called the average) is most likely the measure of central tendency that the reader is most familiar with, but there are others, such as the median, the mode, percentiles, and quartiles.

 The mean, median and mode are all valid measures of central tendency, but under different conditions, some measures of central tendency become more appropriate to use than others.

 A measure of variability is a value that describes the spread or dispersion of a data set to its central value \citep{mccune2009practice}.  If the values of measures of variability are high, it signifies that scores or values in the data set are widely spread out and not tightly centered on the mean. There are three common measures of variability: the range, standard deviation, and variance.

 \textbf{\textit{Mean}}

 The mean (or average) is the most popular and well-known measure of central tendency. It can be used with both discrete and continuous data. An important property of the mean is that it includes every value in the data set as part of the calculation. The mean is equal to the sum of all the values of the variable divided by the number of values in the data set. So, if we have \textit{n} values in a data set and ${(x}_1,\ x_2,\dots ,\ x_n)$ are values of the variable, the sample mean, usually denoted by $\overline{x}$ (denoted by $\mu $, for population mean), is:

\[\overline{x}=\frac{x_1+\ x_2+\dots +x_n}{n}=\frac{\sum^n_{i=1}{x_i}}{n}\]

 Applying this formula to example 2 above, the mean is given by:

\[\overline{x}=\frac{20*6+21*4+22*4+23*2+24*3+25*1}{20}=\frac{435}{20}=21.75\]

 So, the age mean for the 20 individuals is around 22 years (approximately).

 \textbf{\textit{Median}}

 The median is the middle value or the arithmetic average of the two middle values of the variable that has been arranged in order of magnitude. So, 50\% of the observations are greater or equal to the median, and 50\% are less or equal to the median. It should be used with ordinal data. The median (after ordering all values) is as follows:

\[\tilde{x}=\ \left\{ \begin{array}{c}
\frac{x_{\frac{n}{2}}+x_{\frac{n+1}{2}}}{2},\ \mathrm{if}\ \ n\ \mathrm{is\ even} \\ 
\  \\ 
x_{\frac{n+1}{2}},\ \ \mathrm{if}\ n\ \mathrm{is\ odd} \end{array}
\right.\ \]

 In example 2 above, by ordering the age variable values, we have:

20, 20, 20, 20, 20, 20, 21, 21, 21, \textbf{21, 22}, 22, 22, 22, 23, 23, 24, 24, 24, 25

 As \textit{n} is even, the median is the average of the middle values. So $\tilde{x}\mathrm{=}\frac{\mathrm{21+22}}{\mathrm{2}}\mathrm{=21.5}$ is the age median for the sample with 20 individuals.

 \textbf{\textit{Mode}}

 The mode is the most common value (or values) of the variable. A variable in which each data value occurs the same number of times has \textbf{no mode}. If only one value occurs with the greatest frequency, the variable is \textbf{unimodal}; that is, it has one mode. If exactly two values occur with the same frequency, and that is higher than the others, the variable is \textbf{bimodal}; that is, it has two modes. If more than two data values occur with the same frequency, and that is greater than the others, the variable is \textbf{multimodal}; that is, it has more than two modes \citep{mccune2009practice}. The mode should be used only with discrete variables.

 In example 2 above, the most frequent value of age variable is ``20''. It occurs six times. So, ``20'' is the mode of the age variable.

 \textbf{\textit{}}

 \textbf{\textit{}}

 \textbf{\textit{Percentiles and Quartiles}}

 The most common way to report relative standing of a number within a data set is by using percentiles \citep{rumsey2010statistics}. The P${}_{th}$ percentile cuts the data set in two so that approximately P\% of the data is below it and (100$\mathrm{-}$P)\% of the data is above it. So, the percentile of order \textit{p} is calculated by (Mar\^{o}co, 2011):

\[P_p=\left\{ \begin{array}{c}
X_{int(i+1)}\ \ \ \ \ \ \ \ \ \ \ \ \ \ \ \mathrm{if}\ i=\frac{np}{100}\ \ \mathrm{is\ not\ integer} \\ 
\  \\ 
\  \\ 
\frac{X_i+\ X_{i+1}}{2}\ \ \ \ \ \ \ \ \ \mathrm{\ if}\ i=\frac{np}{100}\ \mathrm{is\ integer} \end{array}
\right.\]

 where $n$ is the sample size and $int(i+1)$ is the integer part of $i+1$.

 It is usual to calculate the P${}_{25}$ also called first quartile (Q${}_{1}$), P${}_{50}$ as second quartile (Q${}_{2}$) or median and P75 as the third quartile (Q${}_{3}$).

 In example 2 above, we have:

 20, 20, 20, 20, 20, 20, 21, 21, 21, 21, 22, 22, 22, 22, 23, 23, 24, 24, 24, 25

 Thus, 

\begin{enumerate}
\item  25${}^{th}$ percentile ${(P}_{25})$ or 1${}^{st}$ quartile ($Q_1)$: as $i\mathrm{=}\frac{\mathrm{20*25}}{\mathrm{100}}\mathrm{=}\frac{\mathrm{500}}{\mathrm{100}}\mathrm{=5}$ is integer,
\end{enumerate}

\[P_{25}=Q_1=\ \frac{X_5+\ X_6}{2}=\ \frac{20+20}{2}=20\]

\begin{enumerate}
\item  50${}^{th}$ percentile ${(P}_{50})$ or median: as $i\mathrm{=}\frac{\mathrm{20*50}}{\mathrm{100}}\mathrm{=}\frac{\mathrm{1000}}{\mathrm{100}}\mathrm{=10}$ is integer,
\end{enumerate}

\[P_{50}=Q_2=\tilde{x}=\ \frac{X_{10}+\ X_{11}}{2}=\ \frac{21+22}{2}=21.5\]

\begin{enumerate}
\item  75${}^{th}$ percentile ${(P}_{75})$ or 3${}^{rd}$ quartile ($Q_3)$: as $i\mathrm{=}\frac{\mathrm{20*75}}{\mathrm{100}}\mathrm{=}\frac{\mathrm{1500}}{\mathrm{100}}\mathrm{=15}$ is integer,
\end{enumerate}

\[P_{25}=Q_3=\ \frac{X_{15}+\ X_{16}}{2}=\ \frac{23+23}{2}=23\]

 \textbf{\textit{Range}}

 The range for a data set is the difference between the maximum value (greatest value) and the minimum value (lowest value) in the data set; that is
\[range\ =\ maximum\ value\ -\ minimum\ value\]

 The range should have the same units as those of the data values from which it is computed.

 The interquartile range (IQR) is the difference between the first and third quartiles; that is, $IQR=Q_3-Q_1$ \citep{mccune2009practice}.

 In example 2 above, minimum value=20, maximum value=25. Thus, the range is given by 25-20=5.\textbf{}

 \textbf{\textit{Standard Deviation and Variance}}

 The variance and standard deviation are widely used measures of variability. They provide a measure of the variability of a variable. It measures the offset from the mean of a variable. If there is no variability in a variable, each data value equals the mean, so both the variance and standard deviation for the variable are zero. The greater the distance of the variable' values from the mean, the greater is its variance and standard deviation.

 The relationship between the variance and standard deviation measures is quite simple. The standard deviation (denoted by $\sigma $ for population standard deviation and $s$ for sample standard deviation) is the square root of the variance (denoted by ${\sigma }^2$ for population variance and $s^2$for sample variance).

 The formulas for variance and standard deviation (for population and sample, respectively) are:

\begin{enumerate}
\item  Population variance: ${\sigma }^2=\ \frac{\sum{{(x_i-\mu )}^2}}{N}$, where $x_i$ is the $i$${}^{th}$ data value from the population, $\mu $ is mean of the population, and $N$ is the size of the population

\item  Sample variance:$\ s^2=\ \frac{\sum{{(x_i-\overline{x})}^2}}{n-1}$, where $x_i$ is the $i$${}^{th}$ data value from the sampleSample, $\overline{x}$ is mean of the sample and $n$ is the size of the sample

\item  Population standard deviation: $\sigma =\sqrt{{\sigma }^2}=\ \sqrt{\frac{\sum{{(x_i-\mu )}^2}}{N}}$

\item  Sample standard deviation:$\ s=\sqrt{s^2}=\sqrt{\ \frac{\sum{{(x_i-\overline{x})}^2}}{n-1}}$
\end{enumerate}

\subsection{Charts and Graphs}

 Data can be summarized in a visual way using charts and/or graphs. These are displays that are organized to give a big picture of the data in a flash and to zoom in on a particular result that was found. Depending on the data type, the graphs include pie charts, bar charts, time charts, histograms or boxplots.

 \textbf{}

 \textbf{Pie Charts}

 A pie chart (or a circle chart) is a circular graphic. Each category is represented by a slice of the pie. The area of the slice is proportional to the percentage of responses in the category. The sum of all slices of the pie should be 100\% or close to it (with a bit of round-off error).  The pie chart is used with categorical variables or discrete numerical variables.

 Figure 1 represents the example 1 above.

\begin{figure}[ht]
    \begin{center}
    \includegraphics[scale=2]{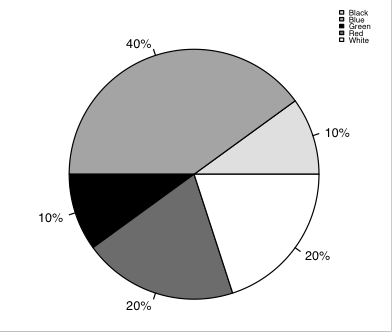}
    \caption{Pie chart example}
\end{center}
\end{figure}

 \textbf{}

 \textbf{Bar Charts}

 A bar chart (or bar graph) is a chart that presents grouped data with rectangular bars with lengths proportional to the values that they represent. The bars can be plotted vertically or horizontally. A vertical bar chart is sometimes called a column bar chart. In general, the x-axis represents categorical variables or discrete numerical variables. 

 Figure 2 and Figure 3 represent the example 1 above.

\begin{figure}
\begin{center}
\includegraphics[scale=2]{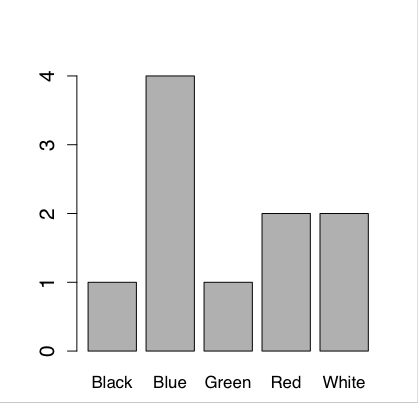}\caption{Bar graph example (with frequencies)}
\end{center}
\end{figure}

\begin{figure}
\begin{center}
\includegraphics[scale=2]{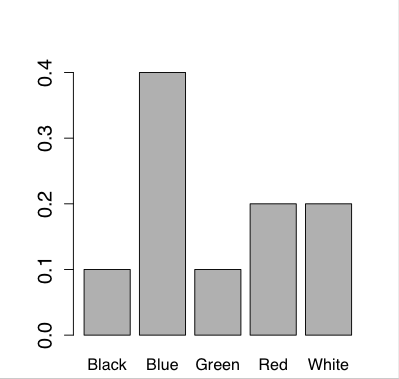}
\caption{Bar graph example (with relative frequencies)}
\end{center}
\end{figure}

\textbf{}

 \textbf{Time Charts}

 A time chart is a data display whose main point is to examine trends over time. Another name for a time chart is a line graph. Typically a time chart has some unit of time on the horizontal axis (year, day, month, and so on) and a measured quantity on the vertical axis (average household income, birth rate, total sales, or others). At each time's period, the amount is shown as a dot, and the dots are connected to form the time chart (Rumsey, 2010).

 Figure 4 is an example of a time chart. It represents the number of accidents, for instance, in a small city along some years.

\begin{figure}
\begin{center}
 \includegraphics*[scale=2]{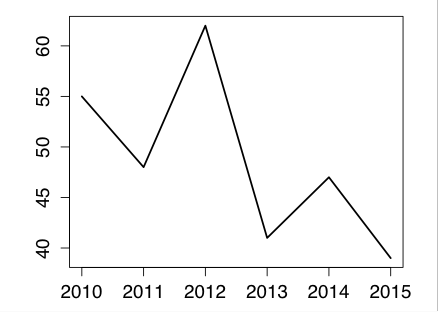}
\caption{Time Chart example}
\end{center}
\end{figure}

 \textbf{}

 \textbf{}

 \textbf{Histogram}

 A histogram is a graphical representation of numerical data distribution. It is an estimate of the probability distribution of a continuous quantitative variable. Because the data is numerical, the categories are ordered from smallest to largest (as opposed to categorical data, such as gender, which has no inherent order to it). To be sure each number falls into exactly one group, the bars on a histogram touch each other but don't overlap (Rumsey, 2010). The height of a bar in a histogram may represent either frequency or a percentage \citep{peers2006statistical}.

 Figure 5 accounts for the histogram of example 3 above.

\begin{figure}
\begin{center}
 \includegraphics*[scale=2]{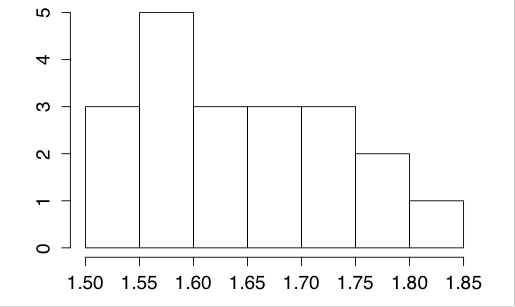}\textbf{}
\caption{Histogram}\textbf{\textit{}}
\end{center}
\end{figure}

 \textbf{Boxplot}

 A boxplot or box plot is a convenient way of graphically depicting groups of numerical data. It is a one-dimensional graph of numerical data based on the five-number summary, which includes the minimum value, the 25${}^{th}$ percentile (also known as Q${}_{1}$), the median, the 75${}^{th}$ percentile (Q${}_{3}$), and the maximum value. In essence, these five descriptive statistics divide the data set into four equal parts (Rumsey, 2010). 

 Some statistical software adds asterisk signs ($*$) or circle signs ($\rm o$) to show numbers in the data set that are considered to be, respectively, outliersOutliers or suspected outliers --- numbers determined to be far enough away from the rest of the data. There are two types of outliers:

\begin{enumerate}
\item  \textit{Outliers} are either 3$\mathrm{\times}$\textit{IQR} or more above the third quartile or 3$\mathrm{\times}$\textit{IQR} or more below the first quartile. 

\item  \textit{Suspected outliersOutliers }are slightly more central versions of outliers: either 1.5$\mathrm{\times}$IQR or more above the third quartile or 1.5$\mathrm{\times}$IQR or more below the first quartile.
\end{enumerate}

 Figure 6 is a boxplot's representation.

 \textbf{}
\begin{figure}
\begin{center}
 \textbf{\includegraphics*[scale=2]{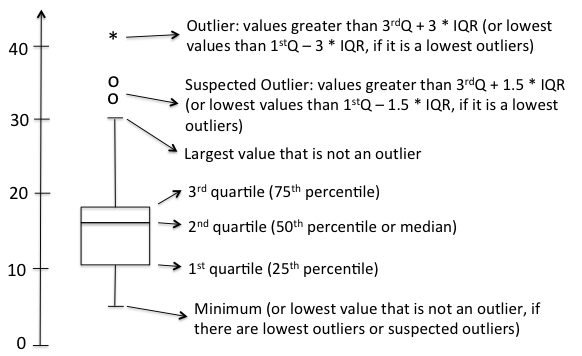}}
\caption{Boxplot}
\end{center}
\end{figure}

 \textbf{}

 \textbf{}

 \textbf{}

\section{Inference Analysis}
 Statistical inference is the process of drawing conclusions about populations or scientific truths from data. This process is divided into two areas: estimation theory and decision theory. The objective of estimation theory is to estimate the value of the theoretical population's parameters by the sample forecasts. The purpose of the decision theory is to establish decisions with the use of hypothesis tests for the population parameters, supported by a concrete measure of the degree of certainty/uncertainty regarding the decision that was taken (Mar\^{o}co, 2011).

\subsection{Inference Distribution Functions (Most Frequent)}

 The statistical inference process requires that the probability density function (a function that gives the probability of each observation in the sample) is known, that is, the sample distribution can be estimated. Thus, the common procedure in statistical analysis is to test whether the observations of the sample are properly fitted by a theoretical distribution. Several statistical tests (e.g., the Kolmogorov-Smirnov test or the Shapiro-Wilk test) can be used to check the sample adjustment distributions for particular theoretical distribution. The following distributions are some probability density functions commonly used in statistical analysis.

\paragraph{Normal distribution}

 The \textbf{normal distribution} or \textbf{Gaussian distribution} is the most important probability density function on statistical inference. The requirement that the sampling distribution is normal is one of the demands of some statistical methodologies with frequent use, called parametric methods (Mar\^{o}co, 2011). A random variable $X$ with a normal distribution of mean $\mu $ and standard deviation $\sigma $ is written as $X\ \sim \ N\ (\mu ,\ \sigma )$. The probability density function (PDF) of this variable is given by:

\begin{table}
\centering
\begin{tabular}{|p{1.2in}|p{0.9in}|} \hline 
$f_X\left(x\right)\mathrm{=}\frac{\mathrm{1}}{\sigma \sqrt{\mathrm{2}\pi }}\mathrm{\ }e^{\mathrm{-}\frac{\mathrm{1}}{\mathrm{2}}\mathrm{\ }{\left(\frac{x\mathrm{-}\mu }{\sigma }\right)}^{\mathrm{2}}}$ & $\boldsymbol{\mathrm{-}}\boldsymbol{\mathrm{\infty }}\boldsymbol{\mathrm{\le }}\boldsymbol{x}\boldsymbol{\mathrm{\ }}\boldsymbol{\mathrm{\le }}\boldsymbol{\mathrm{+}}\boldsymbol{\mathrm{\infty }}$\textbf{} \\ \hline 
\end{tabular}
\end{table}

 The expected value of $X$ is $E\left(X\right)=\mu $, and the variance is $V\left(X\right)={\sigma }^2$. When $\mu =0$ and $\sigma =1$, the distribution is called standard normal distribution and is typically written as $Z\ \sim \ N\ (0,\ 1)$. The letter phi ($\varphi $) is used to denote the standard normal PDF given by:

\begin{table}
\centering
\begin{tabular}{|p{1.1in}|p{0.9in}|} \hline 
$\varphi \left(z\right)\mathrm{=}\frac{\mathrm{1}}{\sqrt{\mathrm{2}\pi }}\mathrm{\ }e^{\mathrm{-}\frac{\mathrm{1}}{\mathrm{2}}\left(z^{\mathrm{2}}\right)}$ & $\mathrm{-}\mathrm{\infty }\mathrm{\le }z\mathrm{\ }\mathrm{\le }\mathrm{+}\mathrm{\infty }$ \\ \hline 
\end{tabular}
\end{table}

 The normal distribution graph has a bell-shaped line (one of the normal distribution names is bell curve) and is completely determined by the mean and standard deviation of the sample. Figure 7 shows a distribution $N\ (0,\ 1)$.

\begin{figure}
\begin{center}
 \includegraphics[scale=2]{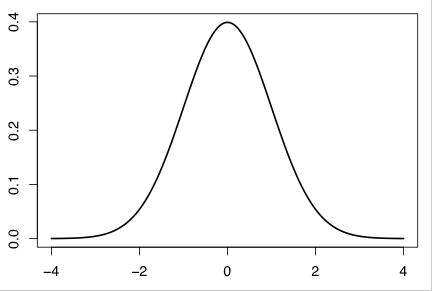}
\caption{Normal distribution}
\end{center}
\end{figure}

 Although there are many normal curves, they all share an important property that allows us to treat them in a uniform fashion. Thus, all normal density curves satisfy the following property, which is often referred to as the Empirical Rule.

\begin{table}
\centering
\begin{tabular}{|p{0.4in}|p{0.6in}|} \hline 
\textbf{RangeRange} & \textbf{Proportion} \\ \hline 
$\mu \pm 1\sigma $ & 68.3 \% \\ \hline 
$\mu \pm 2\sigma $ & 95.5 \% \\ \hline 
$\mu \pm 3\sigma $ & 99.7 \% \\ \hline 
\end{tabular}
\end{table}

 Thus, for a normal distribution, almost all values lie within three standard deviations of the mean.

\paragraph{Chi-Square distribution}

 A random variable $X$ obtained by the sums of squares of $n$ random variables $Z_i\ \sim \ N\ (0,\ 1)$ has a \textbf{chi-square distribution }with $n$ degrees of freedom, denoted as $X^2(n)$. The probability density function (PDF) of this variable is given by \citep{kerns2018introduction}:

\[\mathrm{\ }{\mathrm{f}}_{\mathrm{X}}\left(\mathrm{x}\right)\mathrm{=}\frac{\mathrm{1}}{{\mathrm{2}}^{\frac{\mathrm{n}}{\mathrm{2}}}\mathrm{\ }\mathrm{\bullet }\int^{+\infty }_0{{\mathrm{x}}^{\frac{\mathrm{n}}{\mathrm{2}}\mathrm{-}\mathrm{1}}\mathrm{\ }\mathrm{\bullet }\mathrm{\ }{\mathrm{e}}^{\mathrm{-}\mathrm{X}}\bullet dX}}\mathrm{\ }\mathrm{\bullet }\mathrm{\ }{\mathrm{x}}^{\frac{\mathrm{n}}{\mathrm{2}}\mathrm{-}\mathrm{1}}\mathrm{\ }\mathrm{\bullet }\mathrm{\ }{\mathrm{e}}^{\mathrm{-}\frac{\mathrm{x}}{\mathrm{2}}}\]

 with $\mathrm{n}>0$ e $\mathrm{x}>0$. Figure 8 shows an example of a chi-square distribution.

\begin{figure}
\begin{center}
 \includegraphics[scale=2]{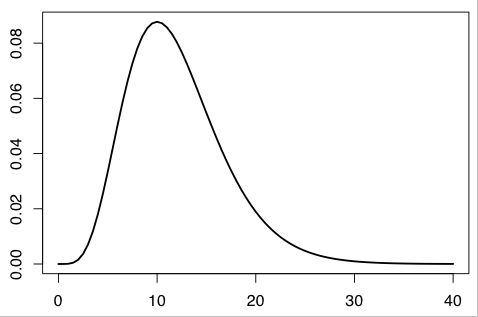}
\caption{Chi-square distribution example}
\end{center}
\end{figure}

 The expected value of $X$ is $E\left(X\right)=n$ and the variance is $V\left(X\right)=2n$. As noted above, the $X^2$distribution is the sum of squares of $n$ variables $N\ (0,\ 1)$. Thus, the central limit theorem (see section \textit{central limit theoremCentral limit theorem}) also ensures that the $X^2$ distribution approaches the normal distribution for high values of $p$.

\paragraph{Student's t-distribution}

 Student's t-distribution is a probability distribution that is used to estimate population parameters when the sample size is small and/or when the population variance is unknown. 

 A random variable $X=\frac{Z}{\sqrt{Y/n}}$ has a student's t-distribution with $n$ degrees of freedom, if $Z\ \sim \ N\ \left(0,\ 1\right)$, and $Y\ \sim \ X^2(n)$ are independent variables. The probability density function (PDF) of this variable is given by (Kerns, 2010):

\[f_X\left(x\right)\mathrm{=}\frac{\tau \left(\frac{n\mathrm{+1}}{\mathrm{2}}\right)}{\sqrt{n\pi }\mathrm{\ }\mathrm{\bullet }\mathrm{\ }\tau \left(\frac{n}{\mathrm{2}}\right)}\mathrm{\ }\mathrm{\bullet }\mathrm{\ }{\left(\mathrm{1+}\frac{x^{\mathrm{2}}}{n}\right)}^{\mathrm{-}\frac{\mathrm{1}}{\mathrm{2}}\left(n\mathrm{+1}\right)}\mathrm{,\ -}\mathrm{\infty }\mathrm{\ }<x\mathrm{<}+\infty \]

 where $\tau \left(\mathrm{u}\right)=\int^{+\infty }_0{{\mathrm{x}}^{\mathrm{u-1}}\mathrm{\ }\mathrm{\bullet }\mathrm{\ }{\mathrm{e}}^{\mathrm{-}\mathrm{x}}}\bullet dX$ and $n>0$. When $n$ increases, this distribution approximates to the centered reduced normal distribution ($N\ (0,\ 1)$). Figure 9 shows an example of a student's t-distribution:

\begin{figure}
\begin{center}
 \includegraphics[scale=2]{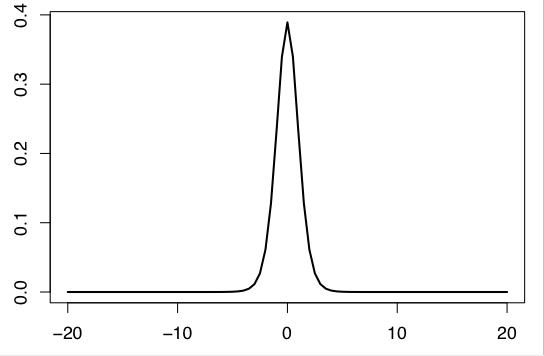}
\caption{Student's t-distribution example}
\end{center}
\end{figure}

 As the centered reduced normal distribution, the student's t-distribution has expected value $E\left(X\right)=0$ and variance $V\left(X\right)=\frac{n}{n-2},\ n>2$.

\paragraph{Snedecor's F-distribution}

 Snedecor's F-distribution is a continuous statistical distribution which arises in the testing of whether two observed samples have the same variance. A random variable $X=\frac{\frac{Y1}{m}}{\frac{Y2}{n}}$ where $Y_1\ \sim \ X^2\left(m\right)$ and $Y_2\ \sim \ X^2\left(n\right)$, has a Snedecor's F-distribution with $m$ and $n$ degrees of freedom, $X\ \sim \ F\ (m,\ n)$. The probability density function (PDF) of this variable is given by (Kerns, 2010):

\[f_X\left(x\right)=\frac{\tau \left(\frac{m+n}{2}\right)}{\tau \left(\frac{m}{2}\right)\ \bullet \ \tau \left(\frac{n}{2}\right)}\ \bullet \ {\left(\frac{m}{n}\right)}^{\frac{m}{2}}\ \bullet x^{\frac{m}{2}-1}\ \bullet \ {\left(1+\frac{m}{n}x\right)}^{-\frac{m+n}{2}},\ x>0\]

 where $\tau \left(\mathrm{u}\right)=\int^{+\infty }_0{{\mathrm{x}}^{\mathrm{u-1}}\mathrm{\ }\mathrm{\bullet }\mathrm{\ }{\mathrm{e}}^{\mathrm{-}\mathrm{x}}}\bullet dX$ and $m>2$ and $n>4$. Figure 10 shows an example of a Snedecor's F-distribution.

\begin{figure}
\begin{center}
 \includegraphics[scale=2]{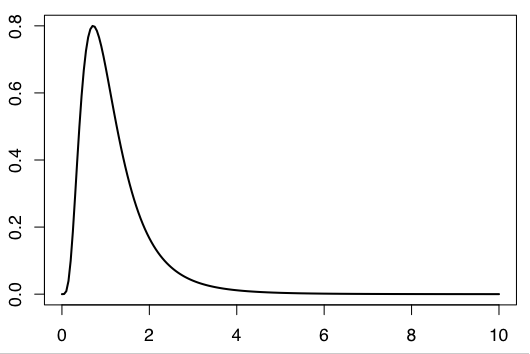}
\caption{Snedecor's F-distribution example}
\end{center}
\end{figure}

 The expected value of $X$ is $E\left(X\right)=\frac{n}{n-2}$ with $n>2$ and the variance is $V\left(X\right)=\frac{2n^2\ \bullet \ (m+n-2)}{m\ \bullet \ {\left(n-2\right)}^2\ \bullet \ (n-4)}$.

\paragraph{Binomial distribution}

 The binomial distribution is the discrete distribution most used in statistical inference to test hypotheses concerning proportions of dichotomous nominal variables (true vs. false, exist vs. non-exists). This distribution is obtained with exactly $n$ successes out of $N$ Bernoulli trials (where the result of each Bernoulli trial is true with probability $p$ and false with probability $q=1-p$). The binomial distribution for the variable $X$ has $n$ and $p$ parameters and is denoted as $X\ \sim \ B\ (n,\ p)$. The probability mass function (PMF) of this variable is given by:

\[f_X\left(x\right)={\left( \begin{array}{c}
n \\ 
x \end{array}
\right)p}^x{(1-p)}^{n-x},\ \ x=0,\ 1,\ 2,\ \dots ,\ n\]

 Figure 11 shows an example of a binomial distribution.

\begin{figure}
\begin{center}
 \includegraphics[scale=2]{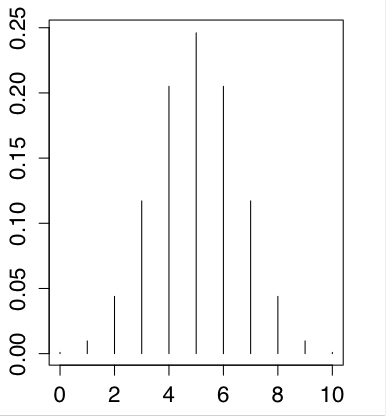}
\caption{Binomial distribution example}
\end{center}
\end{figure}

 The expected value of variable $X$ is $E\left(X\right)=n\bullet p$, and the variance is $V\left(X\right)=n\bullet p\bullet q$. Such as the chi-square distribution or student's t-distribution, the central limit theorem ensures that the binomial distribution is approximated by the normal distribution, when $n$ and $p$ are sufficiently large ($n>20$ and $np>7$; Mar\^{o}co, 2011).

\subsection{Sampling distribution}

 To perform statistical inference - confidence intervals estimation or performing hypothesis testing -- it is necessary to know the distributional properties of the sample, from which it is intended to infer for the theoretical population (Mar\^{o}co, 2011). In the examples given so far, a population was specified, and the sampling distribution of the mean and the range were determined. In practice, the process proceeds the other way: the sample data is collected, and from these data, the parameters of the sampling distribution are estimated.  The mean of a representative sample provides an estimate of the unknown population mean, but intuitively we know that if we took multiple samples from the same population, the estimates would vary from one another. We could, in fact, sample over and over from the same population and compute a mean for each of the samples. In essence, all these sample means constitute yet another "population", and we could graphically display the frequency distribution of the sample means. This is referred to as the sampling distribution of the sample means.

 Some of the sampling distributions commonly used in statistical inference process are presented in the table below (Mar\^{o}co, 2011).

\begin{table}
\centering
\begin{tabular}{|p{0.7in}|p{3.8in}|} \hline 
Statistic & Sampling distribution \\ \hline 
$\overline{X}$ & $\overline{X}\ \sim \ N\left(\mu ,\ \frac{\sigma }{\sqrt{n}}\right)$ if the sampling is with replacement or if the population is too large.\newline \newline $\overline{X}\ \sim \ N\left(\mu ,\ \frac{\sigma }{\sqrt{n}}\times \ \frac{N-n}{N-1}\right)$ if the sampling is without replacement or if the population is small $\left(\frac{n}{N}\right)\le 0.05$.\newline \newline $\frac{\overline{X}-\mu \ }{\frac{S'}{\sqrt{n}}}\sim \ t\left(n-1\right)$ if the population standard deviation is unknown.\newline  \\ \hline 
${S'}^2$ & $\frac{\left(N-1\right){S'}^2}{{\sigma }^2}\ \sim \ \ X^2\left(n-1\right)$ if the variable has normal distribution\newline  \\ \hline 
$\frac{{S'}^2_A}{{S'}^2_B}$ & $\frac{{S'}^2_A}{{S'}^2_B}\sim \ F\left(n_A-1,\ n_B-1\right)$ if the variances have $X^2$ distribution\newline  \\ \hline 
$\hat{P}$ & $\hat{P}\ \sim \ B\left(n,\ p\right)$ for small samples\newline \newline $\frac{\hat{p}-p}{\sqrt{\frac{\hat{p}\left(1-\hat{p}\right)}{n}}}\ \sim \ N\left(0,\ 1\right)$ for large samples (with $n>20$ e $np>\ 5$, where $p$ is the population proportion)  \\ \hline 
\end{tabular}
\end{table}

(Mar\^{o}co, 2011)

 The sample's mean is one of the most relevant statistics for both the theory of estimation as to the theory of decision.

\subsection{Central limit theorem}

 The central limit theorem claims that the distribution of the sample means will be approximately normally distributed if the population has mean $\mu $ and standard deviation $\sigma $, and take sufficiently large random samples from the population with replacement. This will hold true regardless of whether the source population is normal or skewed, provided the sample size is sufficiently large (usually $n>30$). If the population is normal, then the theorem holds true even for samples smaller than 30. In fact, this also holds true even if the population is binomial, provided that $min\left(np,\ n\ \left(1-p\right)\right)>5$, where $n$ is the sample size and $p$ is the probability of success in the population. This means that it is possible to use the normal probability model to quantify uncertainty when making inferences about a population mean based on the sample mean.

 This theorem is particularly useful to justify the use of parametric methods for high dimension samples. When it is not possible to assume that the distribution of the sample mean is normal, particularly when the sample size does not allow the application of the central limit theorem, it is necessary to resort to methods that do not require, in principle, any assumption about the form of the sampling distribution. These methods are referred to generically as non-parametric methods.

\subsection{Hypothesis tests}

 A statistical hypothesis is an assumption about a population parameter. This assumption may or may not be true. Hypothesis tests refer to the formal procedures used by statisticians to accept or reject a statistical hypothesis.

 The best way to determine whether a statistical hypothesis is true would be to examine the entire population. Since that is often impractical, statistical tests are used to determine whether there is enough evidence in a sample of data to infer that a particular condition is true for the entire population. If sample data are not consistent with the statistical hypothesis, the hypothesis is rejected. 

 Hypothesis tests examine two opposing hypotheses about a population: the null hypothesis and the alternative hypothesis: alternative. 

 The null hypothesis, denoted by H${}_{0}$, is the statement being tested. Usually, the null hypothesis is a 

 declaration of the absence of effect or no effect at all and less compromising. The alternative hypothesis: alternative, denoted by H${}_{1}$, is the hypothesis that sample observations are influenced by some non-random cause.

 The H${}_{0}$ should only be rejected if there is enough evidence for a given probability of error or a certain level of confidence, which suggests in fact H${}_{0}$ is not valid.

 However, a hypothesis test can have one of two outcomes: the reader accepts the null hypothesis, or it rejects the null hypothesis. Many statisticians stress with the notion of "accepting the null hypothesis". Instead, they say: you reject the null hypothesis, or you fail to reject the null hypothesis. The distinction between "acceptance" and "failure to reject" is crucial. Whilst acceptance implies that the null hypothesis is true, failure to reject means that the data is not sufficiently persuasive to prefer the alternative hypothesis: alternative to the null hypothesis.

 A hypothesis test is developed in the following steps:

\begin{enumerate}
\item  State the hypotheses. This involves stating the null and alternative hypotheses. The hypotheses are stated in such a way that they are mutually exclusive. That is, if one is true, the other must be false. 

\item  Formulate an analysis plan. The analysis plan describes how to use sample data to evaluate the null hypothesis. The evaluation often focuses around a single test statistic. 

\item  Analyze sample data. Find the value of the test statistic (mean score, proportion, t-score, z-score, etc.) described in the analysis plan. 

\item  Interpret results. Apply the decision rule described in the analysis plan. If the value of the test statistic is unlikely, based on the null hypothesis, reject the null hypothesis. 
\end{enumerate}

 When considering whether the null hypothesis is rejected and the alternative hypothesis is accepted, it is needed to find the direction of the alternative hypothesis statement. This could be a one-tailed test or two-tailed test.

 A \textbf{one-tailed test} is a statistical test in which the critical area of the distribution is one-sided so that it is either greater than or less than a particular value, but not both. If the sample that is being tested falls into the one-sided critical area, the alternative hypothesis will be accepted instead of the null hypothesis. The one-tailed test gets its name from checking the area under one of the tails (sides) of a normal distribution, although the test can be used in other non-normal distributions as well.

 For example, suppose the null hypothesis states that the mean is less than or equal to 10. The alternative hypothesis would be that the mean is greater than 10. The region of rejection would consist of a range of numbers located on the right side of sampling distribution; that is, a set of numbers greater than 10. This represents the implementation of a one-tailed test.

 A \textbf{two-tailed test} is a statistical test in which the critical area of the distribution is two sided and tests whether a sample is either greater than or less than a specified range of values. If the sample that is being tested falls into either of the critical areas, the alternative hypothesis will be accepted instead of the null hypothesis. The two-tailed test gets its name from checking the area under both of the tails (sides) of a normal distribution, although the test can be used in other non-normal distributions.

 For example, suppose the null hypothesis states that the mean is equal to 10. The alternative hypothesis would be that the mean is different to 10, i.e., less than 10 or greater than 10. The region of rejection would consist of a range of numbers located on both sides of sampling distribution; that is, the region of rejection would consist partly of numbers that were less than 10 and partly of numbers that were greater than 10.

\subsection{Decision rules}

 The analysis plan includes decision rules for rejecting the null hypothesis. In practice, statisticians describe these decision rules in two ways - concerning a \textit{p}-value or concerning a region of acceptance.

\paragraph{p-value and statistical errors}

 The \textit{p}-value is the probability of observing a value of the test statistic as extreme or more extreme than the observed test statistic that you computed from the sample. Regarding the distribution associated with the hypothesis test, the \textit{p}-value is calculated as follows:

\begin{enumerate}
\item  For a one-tailed test, the \textit{p}-value is the area to the right (right-tailed test) or left (left-tailed test) of the test statistic.

\item  For a two-tailed test, the \textit{p}-value is two times the area to the right of a positive test statistic or the left of a negative test statistic.
\end{enumerate}

 To make a decision about rejecting or not rejecting H${}_{0}$, it is necessary to determine the cutoff probability for the \textit{p}-value before doing a hypothesis test; this cutoff is called an alpha level ($\alphaup$). Typical values for $\alphaup$ are 0.05 or 0.01. 

 When \textit{p}-value (instead of the test statistic) is used in the decision rule, the rule becomes: If the \textit{p}-value is less than $\alphaup$ (the level of significance), reject H${}_{0}$ and accept H${}_{1}$. Otherwise, fail to reject H${}_{0}$.

 However, incorrect interpretations of \textit{p}-values are very common. The most common mistake is to interpret a \textit{p}-valuep-value as the probability of making an error by rejecting a true null hypothesis (called a type I).

 There are several reasons why \textit{p}-values can't be the error rate. 

 First, \textit{p}-values are calculated based on the assumptions that the null is true for the population and that the difference in the sample is caused entirely by random chance. Consequently, \textit{p}-values can't tell the probability that the null hypothesis is true or false because it is 100\% true from the perspective of the calculations.

 Second, while a small \textit{p}-value indicates that the data are unlikely assuming a true null, it can't evaluate which of two competing cases is more likely: 1) The null is true, but the sample was unusual or; 2) The null is false. Determining which case is more likely requires subject area knowledge and replicate studies.

 For example, supposing that a vaccine study produced a \textit{p}-value of 0.04. The correct way to interpret this value is:\textbf{ }assuming that the vaccine had no effect, it would obtain the observed difference or more in 4\% of studies due to random sampling error. An incorrect way to interpret is: if the null hypothesis is rejected there is a 4\% chance that a mistake is being made.

 \textbf{Types of errors}

 The point of a hypothesis test is to make the correct decision about H0. Unfortunately, hypothesis testing is not a simple matter of being right or wrong. No hypothesis test is 100\% certain because the hypothesis test is based on probability, so there is always a chance that an error has been made. Two types of errors are possible: type I and type II. The risks of these two errors are inversely related and determined by the significance level and the power for the test.

 The following table shows the four possible situations:

\begin{table}[ht]
\centering
\begin{tabular}{|p{0.8in}|p{0.4in}|p{1.4in}|p{1.5in}|} \hline 
\multicolumn{2}{|p{1in}|}{} & \multicolumn{2}{|p{2.8in}|}{Decision} \\ \hline 
\multicolumn{2}{|p{1in}|}{} & Fail to reject & Reject \\ \hline 
Null Hypothesis & True & Correct Decision \newline (probability = 1 - $\alphaup$) & \textbf{Type I} -\newline rejecting the null when it is true (probability = $\alphaup$) \\ \hline 
 & False & \textbf{Type II} - \newline fail to reject the null when it is false (probability = $\betaup$) & Correct Decision \newline (probability = 1 - $\betaup$) \\ \hline 
\end{tabular}
\end{table}

\paragraph{}

 Type I

 When the null hypothesis is true, and it is rejected, it has a type I. The probability of making a type I error is $\alphaup$, which is the significance level set for the hypothesis test. An $\alphaup$ of 0.05 indicates that it is willing to accept a 5\% chance that being wrong when rejecting the null hypothesis. To reduce this risk, a lower value for $\alphaup$ should be used. However, using a lower value for alpha, it will be less likely to detect a true difference if one exists.

 Type II

 When the null hypothesis is false, and it is failed to reject it, it has a type II. The probability of making a type II error is $\betaup$, which depends on the power of the test. It is possible to decrease the risk of committing a type II error by providing that the test has enough power. Ensuring the sample size is large enough to detect a practical difference when one truly exists can do this.

 The probability of rejecting the null hypothesis when it is false is equal to 1--$\betaup$. This value is the power of the test.

 The following example helps to understand the interrelationship between type I, and type II, and to determine which error has more severe consequences for each situation. If there is interest in comparing the effectiveness of two medications, the null and alternative hypotheses are:

 Null hypothesis (H${}_{0}$): $\muup$1= $\muup$2 

 The two medications have equal effectiveness.

 Alternative hypothesis (H${}_{1}$): $\muup$1$\mathrm{\neq}$ $\muup$2 

 The two medications do not have equal effectiveness.

 A type I error occurs if the null hypothesis is rejected, i.e., if it is possible to conclude that the two medications are different when, in fact, they are not. If the medications have the same effectiveness, this error may not be considered too severe because the patients still benefit from the same level of effectiveness regardless of which medicine they take. 

 However, if a type II error occurs, the null hypothesis is not rejected when it should be rejected. That is, it is possible to conclude that the medications have the same effectiveness when, in fact, they are different. This error is potentially life-threatening if the less-effective drug is sold to the public instead of the more effective one.

 When the hypothesis tests are conducted, consider the risks of making type I and type II. If the consequences of making one type of error are more severe or costly than making the other type of error, then choose a level of significance and power for the test that will reflect the relative severity of those consequences.

 \textbf{Acceptance region vs. Rejection region}

 The acceptance region is a range of values. If the test statistic falls within the region of acceptance, the null hypothesis is not rejected. The acceptance region is defined so that the chance of making a type I error is equal to the significance level.

 The set of values outside the acceptance region is called the rejection region. If the test statistic falls within the rejection region, the null hypothesis is rejected. The rejection region is also known as the critical region. The value(s) that separates the critical region from the acceptance region is called the critical value(s). 

 In such cases, we say that the hypothesis has been rejected at the $\alphaup$ level of significance.

\paragraph{Confidence intervals}

 A confidence interval is an estimated range of a parameter of a population. Instead of estimating the parameter by a single value, it is given a range of probable estimates. 

 Confidence intervals are used to indicate the reliability of an estimate. For example, a confidence interval can be used to describe how the results of a search are trustworthy. If all the estimates are equals, a search that results in a small confidence interval is more reliable than one that results in a higher confidence interval. These intervals are usually calculated so that this percentage is 95\%, but it can produce 90\%, 99\%, 99.9\% (or whatever) confidence intervals for the unknown parameter.

 The width of the confidence interval gives some idea of how uncertain the research is about the unknown parameter. A very wide interval may indicate that more data should be collected before anything very definite can be said about the parameter.

 Confidence intervals are more informative than the simple results of hypothesis tests (where we decide "reject H${}_{0}$" or "don't reject H${}_{0}$") since they provide a range of plausible values for the unknown parameter.

 Confidence limits are the lower and upper boundaries/values of a confidence interval, that is, the values that define the range of a confidence interval.

 The upper and lower bounds of a 95\% confidence interval are the 95\% confidence limits. These limits may be taken for other confidence levels, for example, 90\%, 99\%, and 99.9\%.

 The confidence level is the probability value $1-\alpha $ associated with a confidence interval.

 It is often expressed as a percentage. For example, say $\alpha =0.05=5\%$, then the confidence level is equal to $1-0.05=0.95$, i.e. a $95\%$ confidence level. For example, suppose an opinion poll predicted that, if the election were held today, the Conservative party would win 60\% of the vote. The pollster might attach a 95\% confidence level to the interval 60\% plus or minus 3\%. That is, he thinks it very likely that the Conservative party would get between 57\% and 63\% of the total vote.

 \textbf{Summarizing:}

 - A \textit{p}-value is a probability of obtaining an effect as large as or greater than the observed effect, assuming null hypothesis is true

\begin{enumerate}
\item  Provides a measure of strength of evidence against the H${}_{0}$

\item  Does not provide information on magnitude of the effect

\item  Affected by sample size and magnitude of effect: interpret with caution!

\item  Cannot be used in isolation to inform clinical judgment
\end{enumerate}

 - Confidence interval quantifies

\begin{enumerate}
\item  How confident are we about the true value in the source population

\item  Better precision with large sample size

\item  Corresponds to hypothesis testing, but much more informative than \textit{p-}value
\end{enumerate}

 - Keep in mind clinical importance when interpreting statistical significance!

\paragraph{Parametric tests and non-parametric tests}

 During the process of statistical inference, there is often the question about the best hypothesis test for data analysis. In statistics, the test with higher power $(1-\beta )$ is considered the most appropriate and more robust to violations of assumptions or application conditions.

 Hypothesis tests are categorized into two major groups: parametric tests and non-parametric tests. 

 Parametric tests use more information than non-parametric tests and are, therefore, more powerful. However, if a parametric test is wrongly used with data that doesn't satisfy the needed assumptions, it may determine significant differences when truly there isn't one.

 Alternatively, non-parametric tests use less information and, therefore, are more conservative tests than their parametric alternatives. This means that if the reader uses a non-parametric test when he/she has data that satisfies assumptions for a parametric test, the reader can decrease his/her power - i.e. he/she is less likely to get a significant result when, in reality, one exists (significant relationship, significant difference, or other).

\section{Linear Regression Analysis}

 In linear regression model, the functional relationship between the dependent variable and the independent variables $\left(X_i;\ i=1,\ \dots ,\ p\right)$ are \citep{moroco2014analise}:

\[Y_j={\beta }_0+{\beta }_1X_{1j}+{\beta }_2X_{2j}+\dots +{\beta }_pX_{pj}+{\varepsilon }_j,\ \ \ \ \ \ \ \ \ \ \ \ \ \ \ (j=1,\ \dots ,\ n)\]

 In this model, ${\beta }_i$ are the regression coefficientsRegression coefficients and ${\varepsilon }_j$ is the errors or residuals or residuals of the model. ${\beta }_0$ is the y-intercept, and ${\beta }_i$ represents the partial slopes (i.e. a measure of the influence of $X_i$ in $Y$, i.e., the $Y$ variation per variation unit of $X_i$). The term ${\varepsilon }_j$ (errors or residuals of the model) reflects the measurement errors and the natural variation in $Y$. If there is only one independent variable, the model is called simple linear regression. If the model has more than one independent variable, it is called multiple linear regression.

 Considering that the population is not defined, the linear regression analysis should start with the estimation of the regression coefficients from a representative sample of the population under study, using the estimators
\[{\hat{Y}}_j={\widehat{\beta }}_0+{\widehat{\beta }}_1X_{1j}+{\widehat{\beta }}_2X_{2j}+\dots +{\widehat{\beta }}_pX_{pj}\]

 producing sample estimates $b_0,\ b_1,\ \ ...,\ b_p$ of the population's parameters ${\beta }_0,\ {\beta }_1,\ \ ...,\ {\beta }_p$. The commonly used methods for estimation of these coefficients are mostly very laborious. Most software has extensive modules for linear regression. Thus, they eliminate the task of estimating these parameters; its detailed presentation will not be available in this paper.

 \textbf{}

\subsection{Inference In Linear Regression}

 After $b_0,\ b_1,\ \ ...,\ b_p$ estimates being found, the reader should proceed to the evaluation of the quantitative influence of independent variables on the dependent variable in the sample.

\paragraph{ANOVA Statistical Tests}

 The goal now is to evaluate, from sample estimates if, in fact, in the population, some of the independent variables may or may not influence the dependent variable, i.e., if the adjusted model is significant or not (Mar\^{o}co, 2011). This hypothesis could be written as:

\[{\mathrm{H}}_0\mathrm{:}\ {\beta }_1=\ {\beta }_2=\dots ={\beta }_p=0\] 
\[{\mathrm{H}}_{\mathrm{1}}:\ \exists \ i:\ {\beta }_i\neq 0\ \ \ \ \ \ \ (i=1,\ \dots ,\ p)\]

 To test these hypotheses, the total variation in the response $Y$ is divided into two components: one term is the variation in mean response, and the other term is the residual value. This gives the equation 
\[\sum^n_{j=1}{{\left(Y_j-\overline{Y}\right)}^2}=\sum^n_{j=1}{{\left({\hat{Y}}_j-\overline{Y}\right)}^2}+\sum^n_{j=1}{{\left(Y_j-{\hat{Y}}_j\right)}^2}\]

 This equation may also be written as $SST\ =\ SSM\ +\ SSE$, where \textit{SS} is the notation for the sum of squares and \textit{T}, \textit{M}, and \textit{E} are the notation for total, model, and error, respectively.

 For multiple linear regression, the statistic $MSM/MSE$ has an \textit{F} distribution with $\left(DFM,\ DFE\right)=\ \left(p,\ n-p-1\right)$ degrees of freedom, where \textit{p} is the number of independent variables in the model and \textit{n} is the number of variables.

 Hence, if \textit{p}-valuep-value $\mathrm{<}$ $\alpha $, H${}_{0}$ is rejected and it is possible to conclude that, at least one independent variable, has a significant effect on the variance of the dependent variable. This does not mean that the independent variable is the cause of the dependent variable. It can only be said that the adjusted model to the data is significant. However, it should be checked if all or only some independent variables influence the variation of the dependent variable.

\paragraph{Tests on the regression coefficients}

 Similar to the variance analysis for two or more population's averages, by rejecting H${}_{0}$ in the regression ANOVA, we can only conclude that at least one ${\beta }_i$ is significantly different from zero. To determine which of the ${\beta }_i\ \ (i=1,\ \dots ,\ p)$ is nonzero is necessary to carry out multiple ${\beta }_i$ tests. The statistics hypotheses are:
\[{\mathrm{H}}_0\mathrm{:\ }{\beta }_i=\ k\] 
${\mathrm{H}}_{\mathrm{1}}\mathrm{:\ }{\beta }_i\neq \ k\ \ (i=1,\ \dots ,\ p)$ and $k=0$ in most software.

 To test the presented hypotheses, the statistic test has Student's t-distribution with $(n-p-1)$ degrees of freedom. If \textit{p}-value $\mathrm{<}$ $\alpha $, H${}_{0}$ is rejected. Please note that Student's t-test is only valid for each variable, one at a time. The extrapolation of which variables simultaneously influence on the dependent variable is not valid.

\paragraph{Coefficient of determination}

 The coefficient of determination, denoted by R${}^{2}$ or r${}^{2}$ is a number that indicates the proportion of variance in the dependent variable that is predictable from the independent variable.

 It is a statistic used in the context of statistical models whose primary purpose is either the prediction of future outcomes or the testing of hypotheses, by other related information. It provides a measure of how well observed outcomes are replicated by the model, based on the proportion of total variation of the results explained by the model.

 The square of the sample correlation is equal to the ratio of the model squares sum to the total sum of squares: r${}^{2}$ = $SSM/SST$. 

 This formalizes the interpretation of r${}^{2}$ as explaining the fraction of variability in the data explained by the regression model.

 If r${}^{2}$ = 0, the model does not fit the data. If r${}^{2}$ = 1, the adjustment is perfect. The fair r${}^{2}$ value to produce an appropriate adjustment is subjective. In the case of exact sciences, r${}^{2}$ $\mathrm{>}$ 0.9 are accepted as indicators of a good adjustment. However, regarding social sciences, r${}^{2}$ $\mathrm{>}$ 0.5 is acceptable to model's adjustment to the data.

\section{Exploratory Factor Analysis}

  \textbf{}

 Factor analysis is a statistical method used to describe variability among observed, correlated variables.  The goal of performing factor analysis is to search for some unobserved variables called factors. This analysis might lead, for example, to the conclusion that it is possible that three unobserved latent variables are reflected in the variations of seven observed variables. The observed variables are modeled as linear combinations of the possible factors, added the error quantification of this approximation. This added information about the interaction of observed variables could be used for further analysis of the importance of each variable in the context of the dataset.

 Factor analysis is used in many areas of statistical analysis like, for example, marketing, social sciences, psychology and other situations where a reduction of a large set of variables is adequate to the study being provided. This way, some observed variables are substituted by a set of latent variables in a lower amount, and that, therefore, represent the data in a summarized fashion.

 Factor analysis started by being developed before the appearance of modern computers. This beginning of the method was named exploratory factor analysis (EFA). Other variations of factor analysis (for example, confirmatory factor analysis - CFA) will not be explored in this paper. Thus, an example of a factorial analysis is presented below.

 \paragraph{Example}

 Imagine a Ph.D.Ph.D. Supervisor wants to test the hypothesis there are two kinds of students. A student that "procrastinates" his studies, and the student that does "not procrastinate", neither of which is an observed variable. Thus, the supervisor only has access to the grades of the student in the several phases a Ph.D. has. Suppose there are ten stages and the student is classified in all those stages. Additionally, the supervisor has a database of 500 Ph.D. students. By choosing each student randomly from this vast universe of students, imagine the grades as being random variables also. The supervisor hypothesis might clarify that for each of the 10 Ph.D. grades, the score averaged over the group of all students who share some common pair of values for procrastination and "not procrastinating" is some constant multiplied by their level of procrastination plus another constant multiplied by their level of low inertia behaviour, i.e., it is a combination of those two "factors".

 The numbers for a particular stage, by which the two kinds of behavior are multiplied to obtain the expected score, are posited by the hypothesis to be the same for all procrastination level pairs and are called "factor loading" for this subject. For example, the assumption may hold that the average student's aptitude in the field of "State-of-the-Art writing" is $\mathrm{\{}$11 $\mathrm{\times}$ the student's "procrastinating"$\mathrm{\}}$ + $\mathrm{\{}$5 $\mathrm{\times}$ the student's "not procrastinating"$\mathrm{\}}$.

 The numbers 11 and 5 are the factor loadings associated with the task of writing the State-of-the-Art chapter. Other academic tasks may have different factor loadings.

 Two students having similar degrees of procrastination and equal degrees of having low inertia may have different aptitudes in State-of-the-Art writing because individual skills differ from average abilities. That difference is called the "error" - a statistical term that means the amount by which an individual changes from what is average for his or her levels of procrastination.

 The observable data that go into factor analysis would be ten stage's scores of each of the 500 students, a total of 5,000 numbers. The factor loadings and levels of the two kinds of inertia of each student should be inferred from the data.

 \textbf{}

\subsection{The Factor Analysis Model}

 The scores of $p$ population variables, extracted from a population with mean's vector $\mu $ and variance-covariance matrix $\mathrm{\Sigma }$, can be modeled by:
\[x_1={\mu }_1+{\lambda }_{11}f_1+{\lambda }_{12}f_2+\dots +{\lambda }_{1m}f_m+{\eta }_1\] 
\[x_2={\mu }_2+{\lambda }_{21}f_1+{\lambda }_{22}f_2+\dots +{\lambda }_{2m}f_m+{\eta }_2\] 
\[\mathrm{\dots }\] 
\[x_p={\mu }_p+{\lambda }_{p1}f_1+{\lambda }_{p2}f_2+\dots +{\lambda }_{pm}f_m+{\eta }_p\] 
where $f_m$ are factor values (with $m<p$), ${\mathrm{\etaup }}_{\mathrm{p}}$ represent the $p$ specific factors and ${\lambda }_{ij}$ represents the weight of $j$ factor in the variable $i$ (factor loadings), that is, each ${\lambda }_{ij}$ measures the contribution of the $j$ common factor in the variable $i$. Without loss of generality, and for convenience, $x_i$ variables can be centered and reduced as $z_i=(x_i-{\mu }_i)/{\sigma }_i$. Thus, the factor model can be written by:

\[z_i={\lambda }_{i1}f_1+{\lambda }_{i2}f_2+\dots +{\lambda }_{im}f_m+{\eta }_i\ \ \ \ (i=1,\ \dots ,\ p)\]

 Note that ${\mathrm{\lambdaup }}_{\mathrm{ij}}$ values are different depending on whether the analysis is done with the $x_i$ values (factor weights) or $z_i$ (standardized factor weights). It must, therefore, be assumed that \citep{moroco2014analise}:

\begin{enumerate}
\item  Common factors ${(f}_k)$ are independent (orthogonal) and equally distributed with mean 0 and variance 1 $(k=1,\ \dots ,\ m)$.

\item  Specific factors ${(\eta }_j)$ are independent and equally distributed with mean 0 and variance ${\psi }_j,\ \ \left(j=1,\ \dots ,\ p\right)$.

\item  $f_k$ and ${\eta }_j$ are independent.
\end{enumerate}

\subsubsection{Sampling Adequacy}

 Before starting factor analysis, it should be checked whether it is appropriate to the data in the study. For this verification, two methods could be applied: the Bartlett sphericity test and the KMO Measure.

\paragraph{Bartlett Sphericity test}

 Exploratory factor analysis is only useful if the matrix of population correlation is statistically different from the identity matrix. If these are equal, the variables are few interrelated, i.e., the specific factors explain the greater proportion of the variance and the common factors are unimportant. Therefore, it should be defined when the correlations between the original variables are sufficiently high. Thus, the factor analysis is useful in estimation of common factors. With this in mind, the Bartlett Sphericity test can be used. The hypotheses are:

 ${\mathrm{H}}_0\mathrm{:}$ the matrix of population correlations is equal to the identity matrix

 ${\mathrm{H}}_{\mathrm{1}}$: the matrix of population correlations is different from the identity matrix.

 Bartlett's test statistic formula is:

\[X^2=-\left(n-1-\frac{2p+5}{6}\right)\times ln\left|R\right|\]

 where $\left|R\right|$ is the determinant of the correlation matrix, and $\ \left[p\ \times \ \frac{(p-1)}{2}\right]\ $ is the number of degrees of freedom with $p$ as the number of variables. The programming code for this calculation is:

\paragraph{KMO Measure}

 Given the limitations of Bartlett Sphericity test, there are other methods with the same goal that can be used to assess the quality of data. A widely used method is the "measure of the adequacy of sampling Kaiser-Meyer-Olkin" (KMO). KMO checks if it is possible to factorize the primary variables efficiently. But it is based on another idea. 

 The correlation matrix is always the starting point. The variables are more or less correlated, but the others can influence the correlation between the two variables. Hence, with KMO, the partial correlation is used to measure the relation between two variables by removing the effect of the remaining variables.

 The partial correlation matrix can be obtained from the correlation matrix. Considering the inverse of the correlation matrix as $R^{-1}\ =\ (v_{ij})$, the partial correlation as $A\ =\ (a_{ij})$, and the observed correlation matrix as $R=(r_{ij})$, we have:

\[a_{ij}=-\frac{v_{ij}}{\sqrt{v_{ii}\times v_{jj}}}\] 
Thus, the overall KMO index is computed as:

\[KMO=\frac{\sum_i{\sum_{j\neq i}{r^2_{ij}}}}{\sum_i{\sum_{j\neq i}{r^2_{ij}+}}\sum_i{\sum_{j\neq i}{a^2_{ij}}}}\]

 and the KMO index per variable to detect those which are not related to the others is:

\[{KMO}_j=\frac{\sum_{j\neq i}{r^2_{ij}}}{\sum_{j\neq i}{r^2_{ij}}+\sum_{j\neq i}{a^2_{ij}}}\]

 KMO returns values between 0 and 1. A rule of thumb for interpreting the statistic:

\begin{enumerate}
\item  KMO values between 0.8 and 1 indicate the sampling is adequate.

\item  KMO values less than 0.6 indicate the sampling is not appropriate and that remedial action should be taken. Some authors put this value at 0.5, so the researcher should use his judgment for values between 0.5 and 0.6.

\item  KMO Values close to zero means that there are high partial correlations compared to the sum of correlations. In other words, there are widespread correlations, which are a large problem for factor analysis.
\end{enumerate}

 For reference, Kaiser suggested the following classification of the results:

\begin{enumerate}
\item  0 to 0.49 unacceptable

\item  0.50 to 0.59 miserable

\item  0.60 to 0.69 mediocre

\item  0.70 to 0.79 middling

\item  0.80 to 0.89 meritorious.

\item  0.90 to 1.00 marvelous.
\end{enumerate}

\subsection{Retained Factors}

Since it is possible to make a factor analysis to the data in the study, the next step is to find the weights for a set of latent factors. However, this type of mathematical model has multiple possible solutions. This problem is referred to as indeterminacy of Exploratory Factor Analysis (EFA) equation caused by the problem of factors rotation. Therefore, whenever a solution is not interpretable, it can be made a rotation of factors (multiplication by an orthogonal matrix). The “rotation” is equivalent to the translation of factorial axes in the factorial space without changing the orientation of the vectors representing the variables.

\paragraph{Number of factors to be retained }

 Before starting to analyze factors, it is important to know how many factors should be maintained. Several studies were developed to decide this number. \cite{hayton2004factor} state three reasons why this decision is so important. Firstly, it can affect EFA results more than other decisions, such as selecting an extraction method or the factor rotation method, since there is evidence of the relative robustness of EFA with regards to these matters. Secondly, the EFA requires that balance is struck between ``reducing'' and adequately ``representing'' the correlations that exist in a group of variables. Therefore, its very usefulness depends on distinguishing relevant factors from trivial ones. Lastly, an error regarding selecting the number of factors can significantly alter the solution and the interpretation of EFA results. The extraction of fewer factors can lead to the loss of relevant information and a substantial distortion in the solution (for example, in the loading variables). On the other hand, although less problematic, the extraction of an excessive number of factors can lead to factors with a substantial less loading. Thus, it can be difficult to interpret and/or replicate.

 Given the importance of this decision, different methods have been proposed to determine the number of factors to retain.

 \textbf{Kaiser criterion} method suggested by \cite{kaiser1958varimax}. According to his rule, only factors with eigenvalues greater than one are retained for interpretation. Despite the simplicity of this approach, many authors agree that it is problematic and inefficient when it comes to determining the number of factors \citep{ledesma2007determining}. For example, it doesn't make much sense to regard a factor with an eigenvalue of 1.01 as ``major'' and one with an eigenvalue of .99 as ``trivial''. This method should be used together with other methods.

 \textbf{Scree plot:} scree plot proposed by \citep{cattell1966scree}, which involves the visual exploration of a graphical representation of the eigenvalues. In this approach, the eigenvalues are presented in descending order and linked with a line. Afterward, the graph is examined to determine the point at which the last significant drop or break takes place - in other words, where the line levels off. The logic behind this method is that the point divides the critical or major factors from the minor or unimportant factors. Scree plot has been criticized for its subjectivity since there is not an objective definition of the cutoff point between the important and trivial factors. Indeed, some cases may present several drops and possible cutoff points, such that the graph may be ambiguous and difficult to interpret.

 \textbf{Variance explained criteria}  method based on similar conceptual structure is to retain the number of factors that account for a certain percent of extracted variance. The literature varies on how much variance should be explained before the number of factors is sufficient. The majority suggests that 75-90\% of the variance should be accounted. However, some statisticians indicate as much as 50\% of the variance explained is acceptable. As with any criteria method solely depending on variance, this seemingly full standard must be viewed about to the foundational differences between extraction methods. 

To estimate the matrix of factor weights, it is necessary to have an estimate of the communalities. Among the various methods for this estimation, the most popular are Principal Component Analysis, Principal Axis, and Maximum Likelihood Factor Analysis.

\subsection{Principal Component Method}

 The principal component method is based on the determination of the eigenvalues and eigenvectors of the correlation matrix. First, an initial estimate is provided, which is the maximum value of the correlation (i.e., 1). Subsequently, the number of principal components to retain is determined.

\paragraph{Principal Axis Method}

 It is an iterative PCA application to the matrix where communalitiesCommunalities stand on the diagonal in place of 1's. Each iteration refines communalities further until they converge. In doing so, the method seeks to explain variance, not pairwise correlations. Principal Axis method has the advantage in that it can, like PCA, analyze not only correlations but also covariance.

\paragraph{Maximum Likelihood Method}

 Assumes that correlation came from a population having multivariate normal distribution (other methods make no such assumption) and hence the residuals of correlation coefficients must be normally distributed around 0. The loadings are iteratively estimated by ML Communalities approach under the above assumption. The treatment of correlations is weighted by uniqueness values. While other methods just analyze the sample as it is, ML method allows some inference about the population, some fit indices, and confidence intervals are usually computed along with it.

 Succinctly, in most cases, both Principal Component method and Principal Axis method, lead to the same factor structure, and the difference between the methods is mostly conceptual. The Principal Component is the most commonly used method. However, the Principal Axis method is conceptually more attractive, since it assumes a factor structure composed of common factors and specific factors. Consequently, with this method, it is possible to obtain higher factor weights, which facilitates the interpretation of factors. This occurs because they do not have to include the specificity of each variable during the extraction of factors. However, this method is more affected by the indeterminacy of the factors and may cause obtaining very different factor structures from the original data. This disadvantage of the method is particularly penalizing in EFA since it is heavily dependent on the performed sampling. Finally, the maximum likelihood method requires that variables under study present multivariate normal distribution, which is not always easy to validate. This is a reason why the Principal Component method is somewhat recommended. However, it has the advantage that, for large samples, allows the calculation of indices to evaluate the quality of the factor model.

\subsection{Factor Rotations}

 EFA solution is not always interpretable. The factor weights of the variables in common factors can be such that it is not possible to assign a meaning to extracted empirical factors. From the mathematical point of view, the extracted factors are not the only existing ones, and an orthogonal matrix can be multiplied by the matrix of factor weights. The multiplication corresponds to the rotation of the factorial axes and does not alter the communalities or the specific variance, i.e., does not modify the data structure.

 The factorial axes are mathematical structures and not laws of nature. Hence, there is no reason for an axis system to be preferred over another axis system. Moreover, the best axis system is one that produces a factor solution easily interpretable. There are several methods to make the rotation of the factorial axes, including the Varimax method, the Quartimax method and the Oblimin method.

 \textbf{Varimax}

 Varimax, which was developed by Kaiser (1958), is indubitably the most popular rotation method by far. For Varimax, a simple solution means that each factor has a small number of large loadings and a large number of zero (or low) loadings. This simplifies the interpretation because, after a varimax rotation, each original variable tends to be associated with one (or a small number) of factors, and each factor represents only a limited number of variables. Additionally, the factors can often be interpreted from the opposition of few variables with positive loadings to few variables with negative loadings \citep{abdi2003factor}.

 \textbf{Quartimax}

 Quartimax rotation is a form of orthogonal rotation used to transform vectors associated with principal component analysis or factor analysis to a simple structure. It is a particular case of orthomax rotation, which maximizes the sums of squares of the coefficients across the resultant vectors for each of the original variables, Quartimax is opposed to varimax, which maximizes the sums of squares of the coefficients within each of the resultant vectors.

 \textbf{Oblimin}

 Oblimin rotation is a general form for obtaining oblique rotations used to transform vectors associated with principal component analysis or factor analysis to a simple structure. Oblimin is similar to the Orthomax rotation procedures used in orthogonal rotation in that, it too, includes an arbitrary constant used to obtain different rotational properties. While most orthogonal rotations use some form of Orthomax rotation, this is no longer the case with Oblimin rotation for the oblique case.

\subsection{Quality of the factor model}

 Beyond the RMSR mentioned above, a technique widely used in sociology or psychology is the reliability \citep{damasio2012uso}. The reliability of a factor structure may be obtained by several criteria. Among other criteria presented in the literature, the calculation of the level of internal consistencyInternal consistency by Cronbach's alpha $(\alpha )$ is the most used method in cross-sectional studies - when measurements are performed in a single moment \citep{sijtsma2009use}.

 Cronbach's alpha coefficient measures the degree to which the items in an array of data are correlated. Generally, the obtained index varies between 0 and 1. A commonly accepted rule for describing internal consistencyInternal consistency using Cronbach's alpha is:

\begin{enumerate}
\item  0 to 0.49 unacceptable 

\item  0.50 to 0.59 poor 

\item  0.60 to 0.69 questionable 

\item  0.70 to 0.79 acceptable 

\item  0.80 to 0.89 good 

\item  from 0.9 to 1 excellent 
\end{enumerate}

 Cronbach's alpha is influenced by the correlation values of the items and the number of evaluated items. Therefore, factors with a few items tend to have lower Cronbach's alpha and a matrix with high inter-item correlations tend to have a high Cronbach's alpha.

\subsection{Factor Analysis vs. Principal Component Analysis }

 In factor analysis, the different assumption about the communalitiesCommunalities is reflected in a different correlation matrix as compared to the one used in the principal component analysis. Since in principal component analysis all communalities are initially 1, the diagonal of the correlation matrix only contains unities. In factor analysis, the initial communalities are not assumed to be 1. They are estimated (most frequently) by taking the squared multiple correlations of the variables with other variables \citep{rietveld2011statistical}. These estimated communalities are then represented on the diagonal of the correlation matrix, from which the eigenvalues will be determined, and the factors will be retained. After extraction of the factors, new communalities can be calculated, which will be represented in a reproduced correlation matrix \citep{kootstra2004exploratory}.

 The difference between factor analysis and principal component analysis is crucial in interpreting the factor loadings: by squaring the factor loading of a variable, the amount of variance accounted by that variable is obtained. However, in factor analysis, it is already initially assumed that the variables do not account for 100\% of the variance. Thus, as Rietveld \& Van Hout (1993) state, ``although the loading patterns of the factors extracted by the two methods do not differ substantially, their respective amounts of explained variance does!''

\section{Cluster Analysis}

 Cluster analysis was originated in anthropology by \cite{driver1932quantitative} in 1932. It is the task of grouping a set of objects in such a way that objects in the same group or cluster are more similar to each other than to those in other groups or clusters. It is a common technique for statistical data analysis.

 Cluster analysis can be achieved by various algorithms that might differ significantly. Modern notions of clusters include groups with small distances among the cluster members, dense areas of the data space, intervals or particular statistical distributions. Therefore, cluster analysis as such is not a trivial task. It is an interactive multi-objective optimization that involves trial and error. It is also often necessary to modify data preprocessing and model or algorithms parameters until the result achieves the desired characteristics.

 Therefore, in cluster analysis, the clustering of subjects or variables are made from similarity measures or dissimilarity (distance) between two subjects initially, and later between two clusters. These groups can be done using hierarchical or non-hierarchical techniques.

\subsection{Similarity and Dissimilarity }

 The identification of natural clusters of subjects or variables requires that the similarity between these have to be measured explicitly. There are several similarity measures (or proximity) or dissimilarity (or distance) that can be used depending on the variable type (interval, frequency or nominal). In cluster analysis, the most common measures are:

 \textbf{Euclidean distance:} is the distance between two points (p, q) in any dimension of the space and is the most common use of distance. When data is dense or continuous, this is the best proximity measure. Euclidean distance measure is given by:
\[d\left(p,q\right)=\sqrt{\sum^n_{k=1}{{\left(p_k-q_k\right)}^2}}\]

 \textbf{Minkowski distance:} is a metric in a normed vector space, which can be considered a generalization of the Euclidean distance. The Minkowski distance measured between two points (p, q) is given by:
\[d\left(p,q\right)={\left(\sum^n_{k=1}{{\left|p_k-q_k\right|}^c}\right)}^{\frac{1}{c}}\] 
With $c=1$ and $c=2$, the Minkowski metric becomes equal to the Manhattan and Euclidean metrics respectively.

 \textbf{Cosine Similarity:} it is often used when comparing two documents against each other. It measures the angle between two vectors. If the value is zero, the angle between the two vectors is 90 degrees, and they share no terms. If the value is one, the two vectors are the same except for magnitude. Given two vectors of attributes, \textit{u} and \textit{v}, the cosine similarity, ${cos \theta \ }$, is represented as
\[{\mathrm{cos} \theta \ }=\frac{u\bullet v}{\left\|u\right\|\left\|v\right\|}=\ \frac{\sum^n_{i=1}{u_iv_i}}{\sqrt{\sum^n_{i=1}{u^2_i}}\sqrt{\sum^n_{i=1}{v^2_i}}}\] 
where $u_i$ and\textit{ }$v_i$ are components of vector $u$ and $v$, respectively.

 \textbf{Jaccard similarity:} is a standard index for binary variables. It is defined as the quotient between the intersection and the union of the pairwise compared variables between two objects. 

 The Jaccard distance between the objects $i$ and $j$ is given by

\[d\left(i,\ j\right)=\frac{M_{11}}{M_{01}+M_{10}+M_{11}}\] 
\textit{}

 $M_{11}$ represents the total number of attributes where both data objects have a 1; and ${\mathrm{M}}_{\mathrm{10}}\mathrm{,\ }{\mathrm{M}}_{\mathrm{01}}$ represent the total number of attributes where one data object has a 1, and the other has a 0. The total matching attributes are then divided by the total non-matching attributes, plus the matching ones. A perfect similarity score would then be 1.

 \textbf{Similarity measures for variables:} when cluster analysis aims to group variables (and not subjects or items), the appropriate similarity measures are the sample correlation coefficients. In case of continuous variables, Pearson correlation coefficient is the most suitable. For ordinal variables the Spearman correlation coefficient should be used. Finally, for nominal variables, the reader should use the phi coefficient, $\phi =\sqrt{\frac{X^2}{N}}$, where $X^2$ is the chi-square statistic.

\subsection{Hierarchical Clustering}

 Hierarchical techniques appeal to successive steps of aggregation of the considered subjects, individually. Thus, given a set of \textit{N} items to be clustered, and a $N\times N$ distance (or similarity) matrix, the primary process of hierarchical clustering is:

\begin{enumerate}
\item  Start by assigning each item to its cluster, so that if it has \textit{N} items, it now has $N$ clusters, each containing just one item. Let the distances (similarities) between the clusters equal the distances (similarities) between items they contain.

\item  Find the closest (most similar) pair of clusters and merge them into a single cluster, so that now it has one less cluster.

\item  Compute distances (similarities) between the new cluster and each of the old clusters.

\item  Repeat steps 2 and 3 until all items are clustered into a single cluster of size $N$.
\end{enumerate}

 Hierarchical methods of clusters mostly differ in how these distances (in step 3) are calculated. The methods most frequently used are:

\paragraph{Single-linkage clustering}

 Single linkage Clustering (also called connectedness or minimum method) is one of the simplest agglomerative hierarchical clustering methods. In single linkage, the distance between groups is defined as the distance between the closest pair of objects, where only pairs consisting of one object from each group are considered.~

 In single linkage method, $D\ (r,\ s)$ is computed as $D\left(r,\ s\right)=Min\ \left(d\left(i,\ j\right)\right)$, where $i$ is in cluster $r$ and $j$ is in cluster $s$. Thus, the distance between two clusters is given by the value of the shortest link between the clusters.

\paragraph{Complete linkage clustering }

 In complete linkage linkage (also called farthest neighbor), the clustering method is the opposite of single linkage. The distance between groups is defined as the distance between the most distant pair of objects, one from each group.~

 In complete linkage method, $D\ (r,\ s)$ is computed as$\ D\left(r,\ s\right)=Max\ \left(d\left(i,\ j\right)\right)$, where $i$ is in cluster $r$ and object $j$ is in cluster $s$. Thus, the distance between two clusters is given by the value of the longest link between clusters.

\paragraph{Average group linkage}

 With average group linkage, the groups formed are represented by their mean values for each variable (i.e., their mean vector and inter-group distance is defined regarding the distance between two such mean vectors).

 In average group linkage method, the two clusters, $r$, and $s$, are merged such that the average pairwise distance within the newly formed cluster is minimum. Suppose the new cluster formed by combining clusters $r$ and $s$~is labeled as $t$. Then the distance between clusters~$r$ and $s$,~$D\ (r,\ s)$, is computed as $D\ \left(r,\ s\right)=Average\left(d\left(i,\ j\right)\right)$, where observations $i$ and $j$ are in cluster $t$, the cluster formed by merging clusters $r$ and $s$.~

 At each stage of hierarchical clustering, the $r$ and $s$ clusters for which $D\ (r,\ s)$ is minimum, are merged. In this case, those two clusters are merged such that the newly formed cluster, on average, will have minimum pairwise distances between the points.

\paragraph{Average linkage within groups}

 Average linkage within groups is a technique of cluster analysis in which clusters are combined in order to minimize the average distance between all individuals or cases in the resulting cluster. Also, the distance between two clusters is defined as the average distance between all possible pairs of individuals in the cluster that would result if they were combined.

\paragraph{Centroid clustering}

 A cluster centroid is the middle point of a cluster. A centroid is a vector containing one number for each variable, where each number is the mean of a variable for the observations in that cluster.

 The reader can use the centroid as a measure of cluster location. For a particular cluster, the average distance from the centroid is the average of the distances between observations and the centroid. The maximum distance from the centroid is the maximum of these distances.

\paragraph{Ward method Clustering}

 It is an alternative approach for performing cluster analysis. Essentially, it looks at cluster analysis as an analysis of variance problem, instead of using distance metrics or measures of association.

 This method involves an agglomerative clustering algorithm. It will start out at the leaves and work its way to the trunk. It looks for groups of leaves that it forms into branches, the branches into limbs and eventually into the trunk. Ward's method starts out with \textit{n} clusters of size 1 and continues until all the observations are included into one cluster.

 This method is the most appropriate for quantitative variables and not binary variables.

 As there are several available methods, the existence of advantages and disadvantages in using each one of them is visible. Since the ``best'' method of performing hierarchical clustering does not exist, some authors (Mar\^{o}co, 2011) suggest the use of various methods simultaneously. Hence, if all methods produce similarly interpretable solutions, it is possible to conclude that data matrix has natural groupings.

\subsection{Non-hierarchical cluster analysis}

 Non-hierarchical clustering methods are intended in grouping items (and not variables) in a set of clusters whose number is defined a-priori. These methods quickly apply to arrays of large data because it is not necessary to calculate and store a new dissimilarity matrix in each step of the algorithm.

 There are various non-hierarchical methods that differ primarily in the way it unfolds the first aggregation of items in clusters and how the new distances between the centroids of the clusters and the item are calculated. One of the standard methods in most statistical software is the K-means Clustering.

\subsection{K-means Clustering}

 The procedure follows a straightforward and easy way to classify a given data set with a specified number of clusters (assume \textit{k} clusters) fixed a-priori. The main idea is to determine \textit{k} centers, one for each cluster. These centers should be placed in a cunning way because of different location causes a different result. Thus, the better choice is to place them far away from each other, as much as possible. The next step is to take each point belonging to a given data set and associate it to the nearest center. When no point is pending, the first phase is completed, and an early group is done. At this point, the re-calculation of \textit{k} new centroids as barycenter of the clusters resulting from the previous step is done. After this, a new binding has to be done between the same data set points and the nearest new center. A loop has been generated. As a result of this loop, the reader may notice that \textit{k} centers change their location, step by step, until no more changes are done or, in other words, centers do not move anymore.

\section{Classifiers}

Classification is a task of supervised learning, that conveys one or more attributes in order to group sub-populations into different labels or classes. The function that maps the attribution of these classes is called a classifier and its output is therefore discrete. The output is what differentiates classification from the other supervised learning method. Regression output is otherwise continuous, and its mapping function is called estimator.

Knowing the underlying assumptions of the most used classification algorithms may be very useful, when the analyst wants to deepen its data understanding. Prediction accuracy for each problem is usually the most critical feature. However, it may also be useful to have better computational efficiency. Transparency is another issue that may arise because, sometimes algorithms that provide the most accurate models, do not reveal how their models are generated.

In this section, some of the most popular classification algorithms available in the Weka software application - and that will be used in this work - are presented.

The most straightforward algorithm to be tested is the OneR method. “OneR” stands for one rule, i.e., based on a unique attribute. Attributes are ranked based on the training set error rate \citep{holte1993very}. Even considering its simplicity, it can be a relatively accurate method.

Rule sets have several advantages. They are easier to understand and may be used as a first order logic. However, they have some disadvantages too, like poor scaling and noisy data susceptibility. JRip implements the propositional rule learner “Repeated Incremental Pruning to Produce Error Reduction” (RIPPER) proposed by \citep{cohen1995fast}. It is an improved method that grows rules to 100\% accuracy and then prunes over-fitting rules until accuracy starts to decrease.

In a decision tree, each node is either a decision node for an attribute or a leaf node corresponding to a classification. In contrast to rules setting based on error rates, decision trees rely on entropy-based measures (Holte, 1993). The decision stump is merely a one level tree resulting in the same configuration of the OneR algorithm over a single attribute.

Logit Regression has a misleading name and is, in fact, a classification algorithm \citep{bishop2006pattern}. It uses a logistic function to linearly split the outcome in a binary class. Logistic Model Trees (LMT) is a regression tree that uses logistic regression in their decision nodes.

\cite{haykin2010neural} defines an Artificial Neural Network (ANN) as a massively parallel processor, distributed, consisting of simple processing units, which have a natural propensity for storing experiential knowledge and making it available for use.

Sequential Minimal Optimization (SMO) is an optimization of the Support Vector Machines (SVM) algorithm. SVM represents instances as points in space creating clear gaps between different classes \citep{kolvankar2012support}.

Contrary to the previous algorithms, Instance-based learning (IBl) does not record abstractions from instances but predicts based on specific instances \citep{aha1991instance}. This is why it is straightforward to interpret its results.

The other methods that were used in this study are also known as ensemble methods and assume that a combination of classifiers has better results than each classifier isolated. Bagging is an acronym for bootstrap aggregation \citep{breiman1996bagging}. It generates multiple versions of a predictor in order to use them into an aggregated average to predict a class. Boosting “works by sequentially applying a classification algorithm to re-weighted versions of the training data and then taking a weighted majority vote of the sequence of classifiers thus produced” \citep{friedman2000additive}. It improves the performance of any weak learning algorithm by running on various distributions over the training data and combining the resulting models in a composite classifier \citep{freund1996experiments}. Stacking differs from boosting because it combines different learning algorithms and tries to balance their strengths and weaknesses.

\section{Conclusion}

With this document, we present a summary of statistical data analysis concepts. The goal is to present a fast access compilation of most used concepts, when doing statistical data analysis. This might be useful for students, researchers and academics of many areas, where statistical data analysis is used.

\section*{Acknowledgments}
Rui Portocarrero Sarmento also gratefully acknowledges funding from FCT (Portuguese Foundation for Science and Technology) through a Ph.D. grant (SFRH/BD/119108/2016).

\bibliographystyle{apalike}
\bibliography{Report}

\begin{thebibliography}{}

\bibitem[Abdi, 2003]{abdi2003factor}
Abdi, H. (2003).
\newblock Factor rotations in factor analyses.
\newblock {\em Encyclopedia for Research Methods for the Social Sciences. Sage:
  Thousand Oaks, CA}, pages 792--795.

\bibitem[Aha et~al., 1991]{aha1991instance}
Aha, D.~W., Kibler, D., and Albert, M.~K. (1991).
\newblock Instance-based learning algorithms.
\newblock {\em Machine learning}, 6(1):37--66.

\bibitem[Bishop, 2006]{bishop2006pattern}
Bishop, C.~M. (2006).
\newblock {\em Pattern recognition and machine learning}.
\newblock springer.

\bibitem[Breiman, 1996]{breiman1996bagging}
Breiman, L. (1996).
\newblock Bagging predictors.
\newblock {\em Machine learning}, 24(2):123--140.

\bibitem[Cattell, 1966]{cattell1966scree}
Cattell, R.~B. (1966).
\newblock The scree test for the number of factors.
\newblock {\em Multivariate behavioral research}, 1(2):245--276.

\bibitem[Cohen, 1995]{cohen1995fast}
Cohen, W.~W. (1995).
\newblock Fast effective rule induction.
\newblock In {\em Machine learning proceedings 1995}, pages 115--123. Elsevier.

\bibitem[Dam{\'a}sio, 2012]{damasio2012uso}
Dam{\'a}sio, B.~F. (2012).
\newblock Uso da an{\'a}lise fatorial explorat{\'o}ria em psicologia.
\newblock {\em Avalia{\c{c}}ao Psicologica: Interamerican Journal of
  Psychological Assessment}, 11(2):213--228.

\bibitem[Driver and Kroeber, 1932]{driver1932quantitative}
Driver, H. and Kroeber, A. (1932).
\newblock Quantitative expression of cultural relationships. university of
  california publications in american archaeology and ethnology 31: 211--256.
  ester, m.; kriegel, h.-p.; sander, j.; and xu, x. 1996. a density-based
  algorithm for discovering clusters in large spatial databases with noise.
\newblock {\em Driver21131Quantitative Expression of Cultural
  Relationships1932}.

\bibitem[Freund et~al., 1996]{freund1996experiments}
Freund, Y., Schapire, R.~E., et~al. (1996).
\newblock Experiments with a new boosting algorithm.
\newblock In {\em icml}, volume~96, pages 148--156. Citeseer.

\bibitem[Friedman et~al., 2000]{friedman2000additive}
Friedman, J., Hastie, T., Tibshirani, R., et~al. (2000).
\newblock Additive logistic regression: a statistical view of boosting (with
  discussion and a rejoinder by the authors).
\newblock {\em The annals of statistics}, 28(2):337--407.

\bibitem[Haykin, 2010]{haykin2010neural}
Haykin, S. (2010).
\newblock Neural networks: a comprehensive foundation, 1999.
\newblock {\em Mc Millan, New Jersey}, pages 1--24.

\bibitem[Hayton et~al., 2004]{hayton2004factor}
Hayton, J.~C., Allen, D.~G., and Scarpello, V. (2004).
\newblock Factor retention decisions in exploratory factor analysis: A tutorial
  on parallel analysis.
\newblock {\em Organizational research methods}, 7(2):191--205.

\bibitem[Holte, 1993]{holte1993very}
Holte, R.~C. (1993).
\newblock Very simple classification rules perform well on most commonly used
  datasets.
\newblock {\em Machine learning}, 11(1):63--90.

\bibitem[Kaiser, 1958]{kaiser1958varimax}
Kaiser, H.~F. (1958).
\newblock The varimax criterion for analytic rotation in factor analysis.
\newblock {\em Psychometrika}, 23(3):187--200.

\bibitem[Kerns, 2018]{kerns2018introduction}
Kerns, G.~J. (2018).
\newblock Introduction to probability and statistics using r.

\bibitem[Kolvankar et~al., 2012]{kolvankar2012support}
Kolvankar, C., Trivedi, J., Mani, B., Ramanathan, R., and Kadam, S. (2012).
\newblock Support vector machine for learning in artificial intelligence
  systems.
\newblock In {\em Proceedings ofNational Conference on Emerging Trends in
  Engineering \& Technology (VNCET)}, pages 409--412.

\bibitem[Kootstra, 2004]{kootstra2004exploratory}
Kootstra, G.~J. (2004).
\newblock Exploratory factor analysis.
\newblock {\em University of Groningen}.

\bibitem[Ledesma and Valero-Mora, 2007]{ledesma2007determining}
Ledesma, R.~D. and Valero-Mora, P. (2007).
\newblock Determining the number of factors to retain in efa: An easy-to-use
  computer program for carrying out parallel analysis.
\newblock {\em Practical assessment, research \& evaluation}, 12(2):1--11.

\bibitem[McCune, 2009]{mccune2009practice}
McCune, S. (2009).
\newblock {\em Practice makes perfect statistics}.
\newblock McGraw Hill Professional.

\bibitem[Mor{\^o}co, 2014]{moroco2014analise}
Mor{\^o}co, J. (2014).
\newblock An{\'a}lise estat{\'\i}stica com o spss statistics.
\newblock {\em report, 6th ed., Lisboa}.

\bibitem[Peers, 2006]{peers2006statistical}
Peers, I. (2006).
\newblock {\em Statistical analysis for education and psychology researchers:
  Tools for researchers in education and psychology}.
\newblock Routledge.

\bibitem[Rietveld and Van~Hout, 2011]{rietveld2011statistical}
Rietveld, T. and Van~Hout, R. (2011).
\newblock {\em Statistical techniques for the study of language and language
  behaviour}.
\newblock Walter de Gruyter.

\bibitem[Rumsey, 2010]{rumsey2010statistics}
Rumsey, D.~J. (2010).
\newblock {\em Statistics essentials for dummies}.
\newblock John Wiley \& Sons.

\bibitem[Sarmento and Costa, 2017]{sarmento2017comparative}
Sarmento, R. and Costa, V. (2017).
\newblock {\em Comparative Approaches to Using R and Python for Statistical
  Data Analysis}.
\newblock IGI Global.

\bibitem[Sijtsma, 2009]{sijtsma2009use}
Sijtsma, K. (2009).
\newblock On the use, the misuse, and the very limited usefulness of
  cronbach’s alpha.
\newblock {\em Psychometrika}, 74(1):107.

\end{thebibliography}

\end{document}